\documentclass{tlp}
\usepackage{hhline}
\usepackage{amsmath}
\usepackage{amssymb}
\usepackage{latexsym}
\usepackage{hhline}

\newcommand{\dor}{\vee}
\newcommand{\dnf}{~not~}
\newcommand{\dln}{\neg }
\newcommand{\di}{~\leftarrow~}

\newcommand{\dc}{\text{\tt , }}

\newcommand{\ssn}{{\it ssn}}

\newcommand{\lla}{\longleftarrow}
\newcommand{\boxtheorem}{\hfill $\Box$}
\newcommand{\ignore}[1]{}

\newcommand{\nit}[1]{{\it #1}}
\def\allrep{{\cal K}}
\newcommand{\w}{W_{\Pi(r)}}
\newcommand{\W}{{\cal W}_{\Pi(r)}}
\newcommand{\C}{\nit{Core}(\Pi(r))}
\newcommand{\Proof}[1]{\noindent {\em Proof ~of~ #1}\newline}

\newtheorem{definition}{Definition}
\newtheorem{theorem}{Theorem}
\newtheorem{proposition}{Proposition}
\newtheorem{lemma}{Lemma}
\newtheorem{remark}{Remark}
\newtheorem{example}{Example}
\title[Answer Sets for Consistent Query Answers]{{\bf Answer Sets for Consistent Query
Answering in Inconsistent Databases}}

\author[M. Arenas, L. Bertossi and J. Chomicki]{{\bf Marcelo Arenas}\thanks{Current address: University
of Toronto, Department of Computer Science, Toronto, Canada.
marenas@cs.toronto.edu}\\Pontificia Universidad Catolica de Chile,
Departamento de Ciencia de Computacion,\\Santiago, Chile.
marenas@ing.puc.cl \and {\bf Leopoldo Bertossi}\\ Carleton
University, {\it School} of Computer Science, Ottawa, Canada.
bertossi@scs.carleton.ca \and {\bf Jan Chomicki}\\State University
of New York at Buffalo, Dept. of Computer Science and
Engineering,\\Buffalo, NY, USA. chomicki@cse.buffalo.edu}

\begin{document}
\maketitle

\begin{abstract}A relational database is {\em inconsistent} if it does not
satisfy a given set of integrity constraints. Nevertheless, it is
likely that most of the data in it is consistent with the
constraints. In this paper we apply  logic programming based on
answer sets to the problem of retrieving consistent information
from a possibly inconsistent database. Since consistent
information  persists from the original database to every of its
minimal repairs, the approach is based on a specification of
database repairs using {\em disjunctive logic programs with
exceptions}, whose answer set semantics can be represented and
computed by systems that implement stable model semantics. These
programs allow us to declare  persistence  by defaults and
repairing changes by exceptions. We concentrate mainly on logic
programs for binary integrity constraints, among which we find
most of the integrity constraints found in practice.
\end{abstract}

\section{Introduction}\label{sec:intro}

Integrity constraints (IC) capture an important normative aspect
of every database application, whose aim is to guarantee the
consistency of its data. However, it is very difficult, if not
impossible, to always have a consistent database instance.
Databases may become inconsistent with respect to a given set of
integrity constraints. This may happen due, among others, to the
following factors: ~(1) Certain ICs cannot be expressed/maintained
by existing DBMSs. (2) Transient inconsistencies caused by the
inherent non-atomicity of database transactions. (3) Delayed
updates of a datawarehouse. (4) Integration of heterogeneous
databases, in particular with duplicated information. (5)
Inconsistency with respect to \textsl{soft} integrity constraints,
  where transactions in violation of their conditions are not
  prevented from executing.
 (6) Legacy data on which one wants to impose semantic constraints.
 (7) The consistency of the database will be restored by
executing further transactions.
 (8) User constraints than cannot be checked or maintained
 (9) No permission to restore consistency.
 (10) Inconsistent information can be useful.
 (11) Restoring consistency can be a complex and non
deterministic process.

An inconsistent database may be the only source of data, and we
may still want or need to use it. In these cases, one faces the
important problem of characterizing and retrieving the
\textsl{consistent} information from the database. Furthermore,
the database can still provide us with correct answers to certain
queries, making the problem of determining what kinds of queries
and query answers are consistent with the integrity constraints a
worthwhile effort. These problems have been studied in the context
of relational databases and recent publications deal with some of
the issues that arise from trying to define and retrieve
consistent information from an inconsistent relational database.

More specifically, in order to approach the problem of {\em
consistent query answering} (CQA), in \cite{pods99}, a model
theoretic definition of consistent answer to a query posed to an
inconsistent database was introduced. The notion is  based on the
minimal repairs of the inconsistent database: an answer $\bar{t}$
to a first-order query $Q(\bar{x})$ is ~{\em consistent} ~if it is
an answer to the same query in every minimal repair of the
database. A repair must be minimal in the sense that the set of
inserted or deleted tuples (to restore inconsistency) is minimal
under set inclusion.

A possible computational mechanisms for retrieving consistent
answers is as follows: Given a first-order query $Q$ and an
inconsistent database instance $r$, instead of explicitly
computing all the repairs of $r$ and querying all of them, a new
query $T(Q)$ is computed and posed to $r$, the only available
database. The answers to the new query are expected to be the
consistent answers to $Q$. Iterative query transformation
operators were introduced and analyzed with respect to soundness,
completeness and termination in \cite{pods99,celle-bertossi2k}.

Nevertheless, the query rewriting approach has some limitations.
The operators introduced in \cite{pods99,celle-bertossi2k} work
for some particular classes of queries and constraints, e.g. they
cannot be applied to disjunctive or existential queries.

Furthermore, what we have so far is a {\em semantic}, model based
definition of consistent query answer, based on database repairs,
plus a computational mechanism. Nevertheless, we do not have  a
 specification of the database repairs in a logical language of the class of all the
repairs of a given database instance relative to a fixed set of
ICs. Such a description is natural and useful if we  want to
express the explicit or implicit properties that are shared by all
database repairs, e.g. consistent answers, without constructing
those repairs explicitly. Actually, as shown in \cite{icdt01}, it
is easy to find situations where there is an exponential number of
database repairs wrt the original database instance. In
consequence, the specification would be a compact way of
representing the whole class of database repairs.

From such a specification, say ${\it Spec}$, we could: (1) Reason
from ${\it Spec}$. (2) Consistently answer queries by asking:
${\it Spec}~ \models~ Q(\bar{x})?$. (3) Obtain the intended models
of ${\it Spec}$ that should correspond to the database repairs.
(4) Derive algorithms for consistent query answering. (5) Analyze
complexity issues related to consistent query answering. In this
paper, we are motivated mainly by the possibility of retrieving
consistent answers to general first-order queries, extending the
possibilities we developed in \cite{pods99}. We are also
interested in the possibility of obtaining the models of the
specification, i.e. the database repairs. Having the repairs
explicitly available, allows us to analyze different ways  to
restore the consistency of the database. That is, a mechanism for
computing database repairs could be used for conflict resolution.

Notice that consistent answers are non-monotonic in the sense that adding information
to the original database, may cause loosing previous consistent answers. In consequence,
a non-monotonic semantics for ~${\it Spec}$ (or their consequences)  should be
expected.

In this direction, in \cite{annotated} a specification in annotated predicate calculus of the
database repairs was presented. That specification was used to derive algorithms for consistent
query answering and to obtain some complexity results. As expected, the database repairs correspond
to certain minimal models of the specifications. This
approach is based on a non-classical logic.

In the same spirit, in \cite{fqas2k} an alternative specification
of the database repairs was presented. It is based on extended
disjunctive logic programs with an answer sets semantics. The the
database repairs correspond to the intended models of the program.
This paper extends the results presented in  \cite{fqas2k},
addressing several new issues.

Our main contributions are: (1) The introduction and application
of extended disjunctive logic programs with exceptions to the
specification of database repairs and consistent query answering.
(2) A detailed analysis of the correspondence between answer sets
and database repairs for binary integrity constraints. (3) The use
of the specification of database repairs to retrieve consistent
answers to general first order queries. (4) An analysis of the
applicability of the disjunctive well-founded semantics to
consistent query answering. (5) Application of the {\em DLV}
system \cite{E*98} to obtain database repairs and consistent
answers. (6) The use of weak constraints to capture database
repairs based on minimal {\em number} of changes. (7) Extensions
to of the methodology to more general universal constraints and to
referential integrity constraints.

This paper is structured as follows. In section \ref{sec:cqa} we
introduce the notions of database repair and consistent answer to
a query, and the query language. Sections \ref{sec:dlpes}
introduces extended disjunctive logic programs with exceptions. In
section \ref{sec:example} we show a complete, but informal example
that describes the main ideas behind our approach to consistent
query answering and specification of database repairs by means of
logic programs. In section, \ref{sec:bics} we present the repair
programs in their general form for binary integrity constraints.
In section \ref{sec:wf} we analyze the well-founded interpretation
as an approximation to the set of consistent answers, and we
identify cases where it provides the exact solution. In section
\ref{sec:eval}, we discuss how to evaluate the queries. In section
\ref{sec:impl} we show some examples using the {\em DLV} system to
obtain database repairs and consistent answers.  In section
\ref{sec:belief} we shown how database repairs based on minimal
number of changes  can be specified by introducing weak
constraints in our  repair programs. In section \ref{sec:remarks}
we show by means of examples how to extend the methodology to deal
with more general universal constraints and referential integrity
constraints. In section \ref{sec:concl} we discuss related work,
mention some open issues, and draw  conclusions.

\section{Consistent Query Answers}\label{sec:cqa}

A relational database instance $r$ can be represented by  a finite
set of finite relations whose names are part of a database schema.
A database schema can be represented by  a typed language ${\cal
L}$ of first-order predicate logic, that contains a finite set of
sorted predicates and a fixed infinite set of constants $D$. The
language contains a predicate for each database relation and the
constants in $D$ correspond to the elements in the database
domain, that we will also denote by $D$. In consequence, every
database instance has an infinite domain $D$.

The active domain of a database instance $r$ is the set of those
elements of $D$ that explicitly appear in $r$. The active domain
is always finite and we denote it by ${\it Act}(r)$. We may also
have a set of built-in (or evaluable) predicates, like equality,
order relations, arithmetical relations, etc.  In this case, we
have the language ${\cal L}$ possibly extended with these
predicates. In all database instances each of these predicates has
a fixed  and possibly infinite extension. Of course, since we
defined database instances as finite sets of ground atoms, we are
not considering these built-in atoms as members of database
instances.

In addition to the database schema and instances, we may also have
a set of integrity constraints ${\it IC}$ expressed in  a language
${\cal L}$. These are first-order formulas which the database
instances are expected to satisfy.

If a database instance $r$ satisfies ${\it IC}$, what is denoted
by ~$r \models {\it IC}$, we say that it is consistent (wrt ${\it
IC}$), otherwise we say it is inconsistent\footnote{It is natural
to see a relational database instance also as a structure for
interpretation of language ${\cal L}$, in consequence, the
notation ~$r \models {\it IC}$ makes sense.}. In any case, we will
assume from now on  that ${\it IC}$ is  consistent as a set of
first-order sentences.

The original motivation in \cite{pods99} was to consistently
answer first-order queries. We will call them {\em basic queries}
and are defined by the grammar $$B\; ::=\; Atom \;|\; B\wedge B\;
|\; \neg B\; |\; \exists x~B.$$ One way of explicitly asking at
the object level  about the consistent answers to a first-order
query consists in introducing a new logical operator $\allrep$, in
such a way that $\allrep \varphi(\bar{x})$, where
$\varphi(\bar{x})$ is a basic query, asks for the values of
$\bar{x}$ that are consistent answers to $\varphi(\bar{x})$ (or
whether $\varphi$ is consistently true, i.e. true in all repairs,
when $\varphi$ is a sentence). The {\em  $\allrep$-queries} are
similarly defined: $$A \; ::=\; \allrep B\;|\;A\wedge A\;|\;\neg
A\;|\; \exists x. ~A.$$

In this paper we will concentrate mostly on answering {\em basic
$\allrep$-queries} of the form $\allrep B$, where $B$ is a basic
query, but in section \ref{sec:eval} we sketch how to handle
general $\allrep$-queries.

\ignore{we could think of introducing {\em general
$\allrep$-queries} through the grammar: $$C \; ::=\; B\;|\;
\allrep C\;|\;C\wedge C\;|\;\neg C\;|\; \exists x. ~C.$$ }

\begin{definition} \label{def:basic}
 (a) \cite{pods99} Given a database instance $r$ (seen as a set of
ground atomic formulas) and a set of integrity constraints, ${\it
IC}$, a {\em repair}~ of $r$ wrt ${\it IC}$ is
 a database instance $r^\prime$, over the same schema and domain, that satisfies ${\it IC}$ and
 such that $r \Delta r^\prime$, the symmetric  difference of $r$ and $r^\prime$, is minimal under
 set inclusion.\\
(b) \cite{pods99} A tuple $\bar{t}$ is a {\em consistent answer}
to a first-order query $Q(\bar{x})$, or equivalently, an answer to
the query $\allrep Q(\bar{x})$, in a database instance $r$ iff
$\bar{t}$ is an answer to query $Q(\bar{x})$ in every repair
$r^\prime$ of $r$ wrt ${\it IC}$. In symbols: $$r \models ~\allrep
Q[\bar{t}] ~ ~~\Longleftrightarrow~~~ r^\prime \models Q[\bar{t}]
~~\mbox{ for every repair }~ r^\prime ~\mbox{ of }~ r.$$ (c) If
$Q$ is a general $\allrep$-query, then $r \models Q$ is defined as
usual,  (b) being the base case. \boxtheorem
\end{definition}

\begin{example}\label{ex:uno}
Assume there is the following database instance ${\it Salary}$
\begin{center}
\begin{tabular}{c|cc}
$\it{Salary}$ & $\it{Name}$ & $\it{Amount}$\\  \hhline{|-|--|}
 & ${\it V.Smith}$ & 5000\\ & ${\it V.Smith}$ & 8000\\
& ${\it P.Jones}$ & 3000\\ & ${\it M.Stone}$ &
7000\\\hhline{~|--|}
\end{tabular}
\end{center}
\noindent and $\nit{FD}$ is  the functional dependency ~$\it{Name}
\rightarrow \it{Amount}$, meaning that $\it{Name}$ functionally
determines $\it{Amount}$, that is violated by the table
$\it{Salary}$, actually by the tuples with the value
$\it{V.Smith}$ in attribute $\it{Name}$.

The possible repairs are

\begin{center}
\begin{tabular}{c|cccc|cc}
$\it{Salary}_1$ & $\it{Name}$ & $\it{Amount}$ &~~~~~~~~~~~~~&
$\it{Salary}_2$ & $\it{Name}$ &
$\it{Amount}$\\\hhline{|---|~|---|} & $\it{V.Smith}$ & 5000 & & &
$\it{V.Smith}$ & 8000\\ & $\it{P.Jones}$ & 3000 & & &
$\it{P.Jones}$ & 3000\\ & $\it{M.Stone}$ & 7000 & &
&$\it{M.Stone}$ & 7000\\ \hhline{~|--|~~|--|}
\end{tabular}
\end{center}

 \vspace{2mm} Here, $\bar{t}_3 = ({\it P.Jones}, 3000)$ is a consistent answer
to the query ${\it Salary}(\bar{x})$, i.e. ~$r \models \allrep~
{\it Salary}(x,y)[({\it P.Jones}, 3000)]$, but ~$r \not \models
\allrep~ {\it Salary}(x,y)[(\it{V.Smith}, 8000)]$. It also holds~
$r \models \allrep~ ({\it Salary(V.Smith}, 5000) ~\vee ~{\it
Salary(V.Smith}, 8000))$,~ and ~ $r \models \allrep~ \exists X
({\it Salary}(\nit{V.Smith}, X) \wedge X > 4000)$. \boxtheorem\\
\end{example}

Notice that the definition of consistent query answer depends on
our definition of repair. In section \ref{sec:dalal's} we will
consider an alternative definition based on minimal {\em number}
of changes instead of minimal {\em set } of changes.

Computing consistent answer through generation of all possible
repairs is not a natural and feasible alternative \cite{icdt01}.
Instead, an approach based on querying the  available, although
inconsistent, database is much more natural.  In \cite{pods99} a
query rewriting iterative operator $T$ was introduced, that
applied to a given query $Q$ produces a new query $T(Q)$ whose
(ordinary) answers in an instance $r$ are the consistent answers
to $Q$ in $r$.

\begin{example} \label{ex:dos} (example \ref{ex:uno} continued)
The functional dependency ${\it FD}$ can be expressed by means of
the formula  ~$\forall x y z ~(\neg {\it Salary}(x,y) ~\vee~ \neg
{\it Salary}(x,z) ~\vee ~y = z)$. Given the query ~$Q(x,y): ~{\it
Salary}(x,y)$, the consistent answers are: ~$({\it P.Jones},
3000)$, ~$({\it M. Stone},~ 7000)$, but not ~ $({\it V. Smith},~
5000)$, ~$({\it V. Smith},~ 8000)$.

The consistent answers can be obtained by means of the transformed
query ~$T(Q(x,y)) := {\it Salary}(x,y) ~\wedge~ \forall z~(\neg
{\it Salary}(x,z) ~\vee~ y = z)$ posed to the given instance.
\boxtheorem\\
\end{example}

This rewriting approach is not complete for disjunctive or
existential queries,  like ~ $\exists Y~ {\it Salary}({\it
V.Smith},Y)$. We would like to be able to obtain consistent
answers to basic $\allrep$-queries at least.

\section{Logic Programs with Exceptions}
\label{sec:dlpes}

Logic programs with exceptions (LPEs) were first introduced in
\cite{ks91}. They are built with definite extended clauses, that
is, with clauses where the (non-disjunctive) head and the body are
literals (with classical negation) and weak negation (or negation
as failure) may appear in the bodies \cite{glELPb}. Among those
clauses, in a LPE there are positive {\em default} rules, that is
clauses with positive heads, whose conclusions can be overridden
by conclusions derived from {\em exception} rules, which are
clauses with negative heads.

The idea is that exceptions have priority over defaults. To capture this
intuition, a new semantics is introduced, {\em E-answer sets}.

\begin{example}\label{ex:clean}
As an example of LPE, we  present here a program ~$\Pi$~ that
cleans a database instance $r(X,Y)$ from tuples participating in
the violation of the $FD$ ~$X \rightarrow Y$. We start by
introducing a new predicate $r^\prime(X,Y)$ that will store the
tuples in the clean version of the database.

\begin{enumerate}
\item Default rule: ~$r^\prime(X,Y) \lla r(X,Y)$.

It says that every tuple ~$(X,Y)$~ passes from $r$ to $r^\prime$.

\item Negative exception rule: ~$\neg r^\prime(X,Y) \lla r(X,Y),
r(X,Z), {\it not} ~Y = Z$.

It says that tuples $(X,Y)$ where ~$X$~  is associated to  different values
are not accepted in the clean table.

\item Facts: ~the contents of ~$r$ plus ~$X = X \lla$.
\end{enumerate}

Intuitively, rule 2. should have a priority over rule 1.
\boxtheorem\\
\end{example}

The semantics of the program should give an account of the
priorities; they should be reflected in the intended models of the
program.

The semantics is constructed as follows. First, instantiate the
program $\Pi$  in the database domain, making it ground. Now, let
$S$ be a set of ground literals $S = \{L, \ldots\}$. In example
\ref{ex:clean}, $S$ could be something like ~$\{r(a,b), \neg
r(a,b), r^\prime(a,b),$ $ \neg r^\prime(a,b), a = b, \neg a=b,
...\}$. This $S$ is a candidate to be a model, a guess to be
verified, and accepted if properly justified.

Next, generate a new set of ground rules ~$^S\Pi$ according to the
following steps:
\begin{itemize}
\item [(a)] Delete every rule in $\Pi$ containing ~${\it not}~ L$ in the body,
with $L \in S$.
\item [(b)] Delete from the remaining clauses every condition ~${\it
not} ~L$ in the body, when ~$L \notin S$.
\item [(c)] Delete every rule having a positive conclusion $L$ with
~$\neg ~L \in S$.
\end{itemize}

The result is a ground extended logic program without ~${\it
not}$. Now, we say that $S$ is an {\em e-answer set} of the original program if $S$ is the
smallest set of ground literals, such that

\begin{itemize}
\item For any clause ~$L_0 \lla L_1, \cdots, L_m$ ~~in $^S\Pi$, if
$L_1, \cdots, L_m \in S$, then ~$L_0 \in S$.
\item If $S$ contains two complementary literals, then $S$ is the
set of all literals.
\end{itemize}

The e-answer sets are the intended models of the original program.
In example \ref{ex:clean}, the only e-answer set is essentially
the cleaned instance, what is reflected in the existence of only
one e-answer set, where the extension of $r^\prime$ is that of
$r$, but without its conflicting tuples. Notice that this instance
is not a repair, but the intersection of all repairs.

Above, (a) - (b) are as in the {\em answer sets semantics} for
extended logic programs  \cite{glELPb}, but now (c) gives an
account of exceptions.

In order to specify database repairs, we need to extend the LPEs
as presented in \cite{ks91} in order to accommodate also negative
defaults. i.e. defaults with negative conclusions that can be
overridden by positive exceptions, and extended disjunctive
exceptions, i.e. rules of the form $$L_1 \vee \cdots \vee L_k
~\longleftarrow~ L_{k+1}, \ldots, L_r, \nit{not}~ L_{r+1}, \ldots,
\nit{not}~ L_k,$$ where the $L_i$s are literals. In our
application scenario we will need disjunctive exceptions rules,
but  not disjunctive defaults.

The e-answer semantics is extended as follows. The ground program
is pruned  according to a new version of the rule (c) we had in
the previous section:

\begin{itemize}
\item [(c')] Delete every (positive) default having a positive
conclusion $L$, with $\neg L \in S$; and every (negative) default
having a negative conclusion $\neg L$, with $L \in S$.~
\end{itemize}

Applying (a), (b) and (c') to the ground program, we are left with
a  ground disjunctive logic program without $\nit{not}$. If the
candidate set of literals $S$ belongs to $\alpha(^S\Pi)$, the set
of minimal models of program $^S\Pi$, then we say that $S$ is an
{\em e-answer set}.

As described in \cite{ks91} for non-disjunctive programs with
negative exceptions, there is a one to one correspondence between
the e-answer sets of a disjunctive logic program with exceptions
and the answer sets of an extended disjunctive logic program
\cite{glELPb}. It is easy to show a general program transformation
that establishes this correspondence. We will give this
transformation in section \ref{sec:fddbs}, remark
\ref{rem:transf}, but only for the programs we will use. As shown
in \cite{glELPb}, the extended disjunctive program obtained after
the transformation can be transformed in its turn into a
disjunctive normal program (without classical negation) with a
stable model semantics.

\section{Logic Programs for CQA: An Example}
\label{sec:example}

Now, we will use logic programs with exceptions for answering
basic $\allrep$-queries.

Given a set of ICs and an inconsistent database instance $r$, the
first step consists of writing a logic program $\Pi(r)$ having as
e-answer sets the repairs of the original database instance. For
this purpose, we will use the {\em disjunctive logic program with
exceptions} (DLPEs) introduced in section \ref{sec:dlpes}. With
those {\em repair programs} we can specify the class of all the
repairs of a given inconsistent database instance. Next, if a
first-order query is posed with the intention of retrieving all
and only its consistent answers, then a new logic program, a {\em
query program}, that expresses the query, is run together with the
repair program. In this way we can pose and consistently answer
queries we can not handle with the query rewriting approach
presented in \cite{pods99}.

\subsection{The repair program}\label{sec:applLP}

Program $\Pi(r)$  captures the fact that when a database instance
$r$  is repaired most of the data persists, except for some
tuples. In consequence, default rules are introduced: everything
persists from the instance $r$ to the repairs. It is also
necessary to introduce exception rules: everything persists, as
stated by the defaults, unless the ICs are violated and have to be
satisfied.

We illustrate repair programs by means of an example.

\begin{example}\label{ex:exdlpe} Consider an inclusion dependency~
${\it IC}:~ \forall x y~(P(x,y) \rightarrow Q(x,y))$, stating that
every tuple in table $P$ has to be also a tuple in table $Q$, and
the inconsistent database instance $r=\{P(a,b), Q(b,c)\}$. The
repairs of the database are specified by a DLPE ~$\Pi(r)$ obtained
as follows:~ we introduce new predicates $P', Q'$ corresponding to
the repaired versions of the original tables, plus the following
program clauses:

\begin{enumerate}
\item
{\em Facts}:~~ $P(a,b), ~Q(b,c)$.\\

\item
{\em Triggering exception}: ~~~$\neg P'(X,Y) \vee  Q'(X,Y)
~~\leftarrow~~ P(X,Y), ~{\it not}~ Q(X,Y).$\\

 This
rule gives an account of the first two possible steps leading to a
repair of the DB: in order to ``locally" repair the (in this case,
single) IC, either eliminate $(X,Y)$ from $P$ or insert $(X,Y)$
into $Q$. The semantics of these DLPEs
 gives the  disjunction an exclusive
interpretation. We use weak negations in the body of the last rule
in order to give an account of the closed world assumption.\\

\item
{\em Stabilizing exceptions}: ~~~$Q'(X,Y) ~\leftarrow~ P'(X,Y);
~~~~\neg P'(X,Y) \leftarrow \neg Q'(X,Y).$\\

This rule states that eventually the IC has to be satisfied in the
repairs, this kind of exceptions are necessary if there are
interacting ICs and  local repairs alone are not sufficient. The
contrapositive is introduced for technical reasons.\\

\item
{\em Persistence defaults}: ~~~$P'(X,Y) ~\leftarrow~ P(X,Y);~
~~~Q'(X,Y) ~\leftarrow~ Q(X,Y)$; $$\neg P'(X,Y) ~\leftarrow~ ~{\it
not}~ P(X,Y);~~~~ \neg Q'(X,Y) ~\leftarrow~ ~{\it not}~ Q(X,Y).$$
This means that, by default, everything from $r$ is put into
$r^\prime$ and nothing else.

\end{enumerate}

\noindent Rules 2. and 3. have priority over rule 4.\\
\\
It is possible to verify that the e-answer sets of the program are
the expected database repairs: ~~$\{\underline{\neg P'(a,b)}, \neg
Q'(a,b), Q'(b,c), P(a,b), Q(b,c)\};~~~\{P'(a,b),
\underline{Q'(a,b)},$ $ Q'(b,c), P(a,b), Q(b,c)\}$. The underlined
literals represent the deletion of $P(a,b)$ in one repair and the
insertion of $Q(a,b)$, in the other, resp. Notice that $P(a,b)$
and $Q(b,c)$ do not change, because there is no rule able to do
that. \boxtheorem
\end{example}

\subsection{The query program}\label{sec:queryp}

 In order to obtain the consistent answers to a
first-order query $\varphi(\bar{x})$, this query is translated
into a stratified logic program $\Pi(\varphi)$ with new query goal
$\nit{Query}(\bar{x})$ using a standard methodology
\cite{lloyd87}. The predicates appearing in the query program
$\Pi(\varphi)$ will be the repaired, primed versions of the
original database tables, more precisely the set of consistent
answers to $\varphi(\bar{x})$ will be the set $\{\bar{t} ~|~
\nit{Query}(\bar{t}) \in M, \mbox{ for every e-answer set } M
\mbox{ of } \Pi(r) \cup \Pi(\varphi)\}$.

\begin{example} (example \ref{ex:exdlpe} continued) The query
~$\varphi_1(x): ~P(x,a) \vee Q(a,x)$, asking for consistent values
of $x$ in the database instance, can be transformed in the
following query program $\Pi(\varphi_1)$:
\begin{eqnarray*}
Query(X) &\lla& P'(X,a)\\ Query(X) &\lla& Q'(a,X)
\end{eqnarray*}
In order to obtain consistent answers it is necessary to evaluate
the  query goal ~$\nit{Query}(\bar{x})$~ wrt the combined program
~$\Pi(\varphi_1) ~\cup~ \Pi(r)$. Each of the e-answer sets of the
combined program will contain a a set of ground
$\nit{Query}$-atoms. Those $\nit{Query}$-atoms (rather their tuple
arguments) that are present simultaneously in  all the e-answer
sets will be the consistent answers to the original query.

As another example of query, consider ~$\varphi_2(y):~ \exists x
Q(x,y)$. In order to obtain the consistent answers, we keep
$\Pi(r)$, but we run it in combination with the new query program
$\Pi(\varphi_2)$:
\begin{eqnarray*}
\hspace*{4.2cm} Query(Y) &\lla& Q'(X,Y). \hspace{3.8cm} \Box
\end{eqnarray*}

Queries like the ones in the previous example can not be handled
by the rewriting methodology presented in \cite{pods99}.
\end{example}

\ignore{

\begin{example}\label{ex:lloyd} The query ~$Q(Y)$:~
${\it Salary}({\it J. Page},Y) ~\vee~ {\it Salary}({\it V.
Smith},Y)$ can be translated into the program ~$\Pi(Q)$,
consisting of the rules:
\begin{itemize}
\item [] $\nit{Query}(Y) \lla {\it Salary}(X,Y), X = {\it j.Page}$,
\item [] $\nit{Query}(Y) \lla {\it Salary}(X,Y), X = {\it v.Smith}$,
\end{itemize}
where $\nit{Query}$ is the predicate whose extension will contain
the answers to the original query.

If the original query is ~$Q(Y):~ \exists X~ \nit{Salary}(X,Y)$,
the following program can be generated
\begin{itemize}
\item [] $\nit{Query}(Y) \lla {\it Salary}(X,Y)$. \boxtheorem
\end{itemize}
\end{example}

}

\section{DLPEs for Binary Integrity Constraints}\label{sec:bics}

In this section we will introduce the DLPEs for specifying
database repairs, and will give a careful analysis of those
programs for consistent query answering wrt binary integrity
constraints (BICs).

We  represent integrity constraints in the {\em standard format}
\cite{pods99}
\begin{equation}\label{eq:genBics}
\bigvee_{i=1}^n p_i(\bar{x}_i)\vee \bigvee_{i=1}^m \neg
q_i(\bar{y}_i)\vee \varphi,
\end{equation}
where the $p_i, q_i$ are atomic database formulas.
where $\varphi$ is a first-order formula containing only built-in
predicates\footnote{Built-in predicates have a fixed extension in
every database, in particular, in every repair; so they are not
subject to repairs. More details can be found in \cite{pods99}.}
only and whose variables are among the $\bar{x}_i, \bar{y}_i$s;
and there is an implicit universal quantification in front.

Binary integrity constraints are in this standard format, but they
have the restricted  syntactic form
\begin{equation}\label{eq:bic}
\forall \bar{x}~(L_1 \vee L_2 \vee \varphi),
\end{equation}
where $L_1, L_2$ are database literals associated to database
tables, i.e. atomic or negations of atomic formulas whose
predicates are part of the database schema.  That is, in
(\ref{eq:genBics}) the conditions $0 \leq n,m$, $1 \leq n + m \leq
2$ hold.

We represent BICs in this form, as a particular case of the
general standard format, because it is easy to generalize the
program we will give next for BICs to the general case.
Nevertheless, in this paper we will concentrate mainly on BICs.

BICs with one database literal plus possibly a formula containing
built-ins are called {\em unary ICs}. Several interesting classes
of ICs \cite{AbHuVi95} can be represented by BICs: ~ (a) range
constraints, e.g. $P(x,y) \rightarrow x > 5$; (b) non existential
inclusion constraints (example \ref{ex:exdlpe}), (b) functional
dependencies (examples \ref{ex:uno}, \ref{ex:dos}),
 etc. Nevertheless, for referential ICs, like in ~$P(x,y) \rightarrow
 \exists z Q(x,z)$, we need existential
quantifiers or Skolem functions \cite{Fitting:96}. We briefly
consider existential inclusion dependencies in section
\ref{sec:remarks}.

\subsection{Finite domain databases} \label{sec:fddbs}

In this section we will first analyze the case of finite domain
databases. That is, in this section, we will momentarily depart
from our initial assumption that databases have an infinite domain
$D$ (see section \ref{sec:intro}). The reason is that in the
general case, we will be interested in {\em domain independent}
BICs, for which only the active domain is relevant (and finite).

\subsubsection{The change program}\label{sec:change}

In the following, in order to analyze the behavior of DLPEs for
BICs, we will separate the default rules of the programs from the
other rules that represent exceptions. We will concentrate first
on the  program without the defaults, that we will denote by
~$\Pi_\Delta(r)$. This is the part of the program responsible for
the changes.

Splitting the program in this way makes its analysis easier. In
addition, we will see that keeping $\Pi_\Delta(r)$, but using
different form of defaults, we can capture different kinds of
repairs. In this section, we will use defaults (def.
\ref{def:winsprog}) that lead to our notion of repair based on
minimal set of changes (def. \ref{def:basic}). In section
\ref{sec:dalal's}, we will use other defaults that lead to repairs
based on minimal number of changes.

\begin{definition}\label{def:changeprogs}
Given a set of BICs ~${\it IC}$ and an instance $r$, the change
 DLPE, $\Pi_\Delta(r)$, contains the following
rules:

\begin{enumerate}
\item Facts: \begin{enumerate}
\item For every atomic database formula $p(\bar{a})$ such
that $r\models p(\bar{a})$, the fact ~ $p(\bar{a})$.
\item For every $a$ in $D$, the fact ~${\it dom}(a)$.
\end{enumerate}
\item For every IC of the form (\ref{eq:genBics}), the triggering rule
\begin{multline*}
\bigvee_{i=1}^n p^\prime_i(\bar{X}_i) \vee \bigvee_{i=1}^m \neg
q^\prime_i(\bar{Y}_i) \lla\\ {\it dom}(\bar{X}_1,\ldots,
\bar{X}_n) \dc \bigwedge_{i=1}^n \dnf p_i(\bar{X}_i) \dc
\bigwedge_{i=1}^m q_i(\bar{Y}_i) \dc \dnf \varphi.
\end{multline*}
\item For every $1\leq j\leq n$, the stabilizing rule
\begin{multline*}
\bigvee_{i=1}^{j-1} p^\prime_i(\bar{X}_i) \vee \bigvee_{i=j+1}^n
p^\prime_i(\bar{X}_i) \vee \bigvee_{i=1}^m \neg
q^\prime_i(\bar{Y}_i) \lla\\ {\it dom}(\bar{X}_1,\ldots,
\bar{X}_n, \bar{Y}_1, \ldots, \bar{Y}_m) \dc \neg
p^\prime_j(\bar{X}_j) \dc \dnf \varphi.
\end{multline*}
For every $1\leq j\leq m$, the stabilizing rule
\begin{multline*}
\bigvee_{i=1}^n p^\prime_i(\bar{X}_i) \vee \bigvee_{i=1}^{j-1}
\neg q^\prime_i(\bar{Y}_i) \vee \bigvee_{i=j+1}^m \neg
q^\prime_i(\bar{Y}_i) \lla\\ {\it dom}(\bar{X}_1,\ldots,
\bar{X}_n, \bar{Y}_1, \ldots, \bar{Y}_m) \dc q^\prime_j(\bar{Y}_j)
\dc \dnf \varphi.
\end{multline*}
\end{enumerate}
In these rules, ${\it dom}(\bar{X}_1,\ldots,\bar{X}_n)$ is an
abbreviation for the conjunction of cases of membership to ${\it
dom}$ of all the components in the $\bar{X}_i$s. Of course,
depending on the syntax, it may be necessary to unfold the formula
$\varphi$ appearing in the bodies in additional program rules, but
$\varphi$ will usually be a conjunction of literals.\boxtheorem\\
\end{definition}

Notice that the sets of rules 2. and 3. in example \ref{ex:exdlpe}
have the form of  these general triggering and stabilizing rules,
respectively.

It is always the case that for BICs, the stabilizing rules in
$\Pi_\Delta(r)$ do not contain disjunctions in the heads. This can
also be seen in example \ref{ex:exdlpe}. Only triggering rules are
properly disjunctive.

\newpage

\begin{definition}\label{def:model}
A {\em model} of a DLPE, $\Pi$, is a set of ground literals, $S$,
that does not contain complementary literals and satisfies $\Pi$
in the usual logical sense, but with weak negation interpreted as
not being an element of $S$. \boxtheorem
\end{definition}

\begin{definition}
Given a model $S$ of $\Pi_\Delta(r)$, we define the database
instance corresponding to $S$ by

\hspace*{1cm} $I(S)                                   =
\{p(\bar{a}) \mid p^\prime(\bar{a}) \in S\} ~\cup~ \{ p(\bar{a})
\mid p(\bar{a})\in S \text{ and } \neg p^\prime(\bar{a}) \notin
S)\}$. \boxtheorem\\
\end{definition}

Notice that, for a given model $S$ of the change program, $I(S)$
merges in one new instance of the schema all the positive primed
tuples with all the old, non primed tuples that persisted, i.e.
that their negative primed version do not belong to the model.
Since we do not have persistence defaults in $\Pi_\Delta(r)$,
persistence is captured and imposed through $I(S)$, by keeping in
it all the atoms from the original database that were not
discarded via the primed predicates.

\begin{proposition}\label{l1}
Given a database instance $r$ and a set of BICs ${\it IC}$, if $S$
is a model of $\Pi_\Delta(r)$, then $I(S)$ satisfies ${\it IC}$.
\boxtheorem\\
\end{proposition}

\begin{definition}
Given database instances $r$ and $r^\prime$ over the same
schema and domain, we define
\begin{eqnarray*}
S(r,r^\prime) &= & \{p(\bar{a}) \mid r\models p(\bar{a})\}~ \cup~
\{ p^\prime(\bar{a})\mid r^\prime\models p(\bar{a})\} ~\cup~
\{\neg p^\prime(\bar{a})\mid r^\prime \not\models p(\bar{a})\} \\
&&\cup ~\{{\it dom(a)}\mid a\in D\}. \hspace{7.2cm}\Box
\end{eqnarray*}
\end{definition}

$S(r,r^\prime)$ collects the maximal set of literals that can be
obtained from two database instances. It contains everything from
both $r$ and $r^\prime$. The atoms corresponding to the second
argument are primed. Negative literals corresponding to the first
argument, intended to be the original databases instance, are not
considered, because we will apply weak negation to them.

\begin{proposition}\label{l2}
Given a database instance $r$ and a set of BICs $IC$, if
~$r^\prime$ satisfies ${\it IC}$, then  $S(r,r^\prime)$ is a model
of $\Pi_\Delta(r)$.\boxtheorem\\
\end{proposition}

In the following we will be considering subsets of $S(r,r')$. The
previous result tells us that its subsets can be potential models
of the program. $S(r,r')$ can be a large model, in the sense that
the difference between $r$ and $r'$ may not be minimal.

\begin{proposition}\label{exstable}
For BICs, the change program $\Pi_\Delta(r)$ has an answer set.
\boxtheorem\\
\end{proposition}

\subsubsection{The repair program}\label{sec:repprog}

Program $\Pi_\Delta(r)$ gives an account of the changes in the
original instance that are needed to produce the repairs, but the
actual repairs contain data that persists from the original
instance. This can be captured by adding persistence defaults.

\begin{definition} \label{def:winsprog}
The repair program $\Pi(r)$ consists of the rules in program
$\Pi_\Delta(r)$ (def. \ref{def:changeprogs}) plus  the following
two rules for each predicate $p$ in the original database:

\begin{itemize}
\item [4.] Persistence {\em defaults}
\begin{eqnarray*}
p^\prime(\bar{X}) &\lla& p(\bar{X})\\
\hspace*{2.3cm} \dln
p^\prime(\bar{X}) &\lla& {\it dom}(\bar{X}) \dc
p(\bar{X}). \hspace{4.3cm} \Box
\end{eqnarray*}
\end{itemize}
\end{definition}

\begin{remark} \label{rem:transf}
As shown in \cite{ks91}, the DLPE 1.-- 4., which has an
e-answer semantics, can be transformed into a disjunctive extended
logic program with answer set semantics, by transforming the
persistence defaults into

\begin{itemize}
\item [4'.] Persistence {\em rules}
\begin{eqnarray*}
p^\prime(\bar{X}) &\lla& p(\bar{X}) \dc \dnf \dln
p^\prime(\bar{X})\\
\neg p^\prime(\bar{X}) &\lla& {\it dom}(\bar{X}) \dc
\dnf p(\bar{X}) \dc \dnf p^\prime(\bar{X}).
\end{eqnarray*}
\end{itemize}
As shown in \cite{glELPb}, the resulting extended disjunctive
normal program can be further transformed into a disjunctive
normal program with a stable model semantics, with a one to one
correspondence between answer sets and stable models. For this
reason, we will interchangeably use the terms answer sets and
stable models. \boxtheorem
\end{remark}

\begin{proposition}\label{aux2}
Given a database instance $r$ over a finite domain, and a set of
BICs $\nit{IC}$, if $ S_M$ is an answer set\footnote{In the
programs we are considering so far, namely $\Pi_\Delta(r)$ and
$\Pi(r)$ with persistence rules instead of persistence defaults,
we do not find any defaults. In consequence, we can talk about
answer sets as in \cite{glELPb} instead of e-answer sets
\cite{ks91}.} of $\Pi_\Delta(r)$, then
\begin{multline}\label{e2}
S =  S_M \cup \{p^\prime(\bar{a}) \mid p(\bar{a})\in S_M \text{
and }  \dln p^\prime(\bar{a})\not\in S_M\}\cup\\ \{ \dln
p^\prime(\bar{a}) \mid p(\bar{a})\not\in S_M \text{ and }
p^\prime(\bar{a})\not\in S_M\}
\end{multline}
is an answer set of $\Pi(r)$. \boxtheorem\\
\end{proposition}

In order to establish the correspondence between the answer sets
of the repair program $\Pi(r)$ and the repairs of $r$, we need the
following lemma. It says that whenever we build an answer set $S$
with literals taken from $S(r,r^\prime)$, and $r'$ satisfies the
ICs and is already as close as possible to $r$, then in $S$ we
 can get essentially $r'$ only.  The condition that $S$ is contained in
$S(r,r^\prime)$ makes sure that the literals in $S$ are taken from
the right, maximal set of literals.

\begin{lemma}\label{l3}
Let $r$ and $r^\prime$ be database instances  over the same schema
and domain, and ${\it IC}$, a set of BICs. Assume that ~$r^\prime
\models {\it IC}$ and the symmetric difference
$\Delta(r,r^\prime)$  is a minimal element under set inclusion in
the set ~$\{\Delta(r,r^*)\mid r^*\models IC\}$. Then,  for every
answer set $S$ of $\Pi_\Delta(r)$ contained in $S(r,r^\prime)$, it
holds ~$r^\prime=I(S)$. \boxtheorem\\
\end{lemma}

\ignore{
 By proposition \ref{l3}, we know that for BICs,  the set
of minimal models of $\Pi_\Delta(r)$, among which we find the
answer sets of the program, will contain all repairs. }

\begin{theorem}\label{tw}
If $\Pi(r)$ is the program $\Pi_\Delta(r)$ plus rules 4'., for a
finite domain database instance $r$ and a set of BICs ${\it IC}$,
then it holds:

\begin{enumerate}
\item For every repair $r^\prime$ of $r$ wrt ${\it IC}$, there exists an
answer set $S$ of $\Pi(r)$ such that $r^\prime = \{p(a) ~|~ p'(a)
\in S\}$.
\item For every answer set $S$ of $\Pi(r)$, there exists a
repair $r^\prime$ of $r$ wrt ${\it IC}$ such that $r^\prime =
\{p(a) ~|~ p'(a) \in S\} $. \boxtheorem\\
\end{enumerate}
\end{theorem}

 In the case of finite domain databases, the domain can be
and has been declared. In this situation, we can handle any set of
binary ICs, without caring about their safeness or domain
independence.

\begin{example} \label{ex:finite}
Let us take ~$D=\{a,b,c\}$, $r=\{p(a)\}$ and ${\it IC}=\{\forall x
p(x)\}$. In this case, the program $\Pi(r)$ is
\begin{eqnarray*}
p^\prime(X) &\lla& p(X), {\it not}  ~\neg p^\prime(X)\\ \neg
p^\prime(X) &\lla& {\it dom}(X), {\it not}~ p(X), {\it not}~
p^\prime(X)\\ p^\prime(X) &\lla& {\it dom}(X), {\it not}~ p(X)\\
p^\prime(X) &\lla& {\it dom}(X)\\ {\it dom}(a) &\lla&\\
 {\it dom}(b) &\lla&\\
 {\it dom}(c) &\lla&\\
 p(a) &\lla&
\end{eqnarray*}
The only answer set is ~$\{{\it dom}(a), {\it dom}(b), {\it
dom}(c), p(a), p^\prime(a), p^\prime(b), p^\prime(c)\}$, that
corresponds to the only repair~ $r^\prime = \{p(a), p(b), p(c)\}$.
\boxtheorem\\
\end{example}

In this example, the IC demands that every element in  domain $D$
belongs to table $p$; and this is possible to satisfy because the domain is finite.
Nevertheless, if the domain $D$ were infinite this would not be
possible, because relational tables contain finitely
many tuples. So, this kind of ICs cannot be handled  in infinite domains.

\subsection{Infinite domain databases}\label{sec:inf}

Now we consider only ICs that are {\em domain independent}
\cite{ullmanI}.  For these ICs only the active domain matters. In
particular, checking their satisfaction in an instance $r$ can be
done considering the elements of ${\it Act}(r)$ only. The IC in
example \ref{ex:finite} is not domain independent.

For domain independent BICs all previous lemmas and theorems still
hold if we have infinite database domains $D$. To obtain them, all
we need to do is to use a predicate ~${\it act}_r(x)$, standing
for the active domain ${\it Act}(r)$ of instance $r$, instead of
predicate ${\it dom}(x)$. This is because, for domain independent
BICs, the database domain can be considered to be the finite
domain ${\it Act}(r)$. Furthermore, in this case we can omit the
$\nit{dom}$ facts in 1. of  $\Pi_\Delta(r)$ (definition
\ref{def:changeprogs}). In consequence, we have the following
theorem, first stated in \cite{fqas2k}.

\begin{theorem} \label{theo:domainindep}  For a set of domain independent
binary integrity constraints and a database instance $r$, there is
a one to one correspondence between the answers sets of the repair
program $\Pi(r)$ and the repairs of $r$. \hfill $\Box$
\end{theorem}

As a consequence of this general result, we have that our DLPEs
$\Pi(r)$ correctly specify the repairs of relational databases
that violate usual integrity constraints like range constraints,
key constraints, functional dependencies, and non-existential
inclusion dependencies.

\section{Well-Founded Consistent Answers} \label{sec:wf}

Computing the stable model semantics for disjunctive programs is
$\Pi^P_2$-complete in the size of the ground program\footnote{See
\cite{voronkov} for a review of complexity results in logic
programming.}. In some cases, computing  consistent answers can be
done more efficiently.

The intersection of all answer sets of a extended disjunctive
logic program contains the well-founded interpretation for such
programs  \cite{infComp}, that can be computed in polynomial time
in the size of the ground program. This interpretation may be
partial and not necessarily a model of the program. Actually, it
is a total interpretation if and only if it is the only answer
set.

The well-founded interpretation, $W_{\Pi(r)} = <W^+,W^->$, of
program $\Pi(r)$, where $W^+,W^-,W^u$ are the sets of true
positive, negative, unknown literals, resp., is the given by the
fixpoint ${\cal W}_{\Pi(r)}^\omega(\emptyset)$ of operator ${\cal
W}_{\Pi(r)}$, that maps interpretations to interpretations
\cite{infComp}.
 More precisely, assuming that we
have the ground instantiation of the repair program $\Pi(r)$,
 ${\cal W}_{\Pi(r)}(I)$ is defined on interpretations
$I$ that are sets of ground literals (without pairs of
complementary literals) by: ~${\cal W}_{\Pi(r)}(I) :=
T_{\Pi(r)}(I) \cup \neg. \nit{GUS}_{\Pi(r)}(I)$.

Intuitively, $T_{\Pi(r)}$ is the immediate consequence operator
that declares a literal true whenever there is ground rule
containing it in the head, the body is true in $I$ and the other
literals in the (disjunctive) head are false in $I$.
$\neg.\nit{GUS}_{\Pi(r)}(I)$ denotes the set of complements of the
literals in  $\nit{GUS}_{\Pi(r)}(I)$, being the latter the largest
set of unfounded literals, those that definitely cannot be derived
from the program and the set $I$ of assumptions; in consequence
their complements are declared true. The well-founded
interpretation, $W_{\Pi(r)}$, is the least fixpoint ${\cal
W}_{\Pi(r)}^\omega(\emptyset)$ of ${\cal W}_{\Pi(r)}$. More
details can be found in \cite{infComp}.

The intersection of all answer sets of $\Pi(r)$ is
$$\nit{Core}(\Pi(r)) := \bigcap \{S ~|~ S \mbox{ is an answer set
of } \Pi(r)\}.$$ Interpretation $W_{\Pi}$, being  a subset of
$\nit{Core}(\Pi(r))$, can be used as an approximation from below
to the core  that can be computed more efficiently than all
database repairs, or their intersection in the general case.
Nevertheless, it is possible to identify classes of ICs for which
$W_\Pi(r)$ coincides with $\nit{Core}(\Pi(r))$. In these cases,
the core is no longer approximated by $W_{\Pi(r)}$, but computed
exactly.

In order to prove these results, we will assume, as in section
\ref{sec:fddbs}, that we have a finite database domain $D$. We
know that the results obtained under this hypothesis still hold
for infinite domain databases and domain independent integrity
constraints. In consequence, program $\Pi(r)$ contains domain
predicates. The domain facts belong to every answer set and are
obtained after the first iteration of the well-founded operator.

\begin{proposition} \label{prop:wfs}
For a database instance $r$,  and a set of ICs containing
functional dependencies and unary ICs only\footnote{Remember that
they are BICs with at most one database literal in the standard
format (\ref{eq:bic}), plus built-ins. They include range
constraints, e.g. ~$stock(x) \rightarrow 100\leq x$, stating that
products in stock may not go below 100 units.}, the
$\nit{Core}(\Pi(r))$ of program $\Pi(r)$ coincides with
$W_{\Pi(r)}$, the well-founded interpretation of program $\Pi(r)$.
\boxtheorem\\
\end{proposition}

As corollary of this proposition and results presented in
\cite{infComp} about the computational complexity of the
disjunctive well-founded interpretation, we obtain that, for FDs
and unary constraints, $\nit{Core}(\Pi(r))$ can be computed in
polynomial time in the size of the ground instantiation of
$\Pi(r)$, a result first established in \cite{icdt01} for FDs. In
particular, we can consistently answer non-existential conjunctive
queries in polynomial time, because we can use
$\nit{Core}(\Pi(r))$ only. Furthermore, in \cite{icdt01}, for the
case of functional dependencies, some conditions on queries are
identified under which one can take advantage of computations on
the core to answer aggregate queries more efficiently.

For programs of the kind we may have for BICs, it is not always
the case that the core coincides with the well-founded
interpretation.

\begin{example} \label{ex:noncore}
Consider the BICs ~$IC = \{q \vee r, ~s \vee \neg q, ~s \vee \neg
r\}$ and the empty database instance. The program $\Pi(r)$ wrt
$IC$ is\\

\noindent {\em Triggering rules}:
\begin{eqnarray*}
~~~ q' \vee r' &\lla& not~ q, not~ r\\ s' \vee \neg q' &\lla& not~
s,q\\ s' \vee \neg r' &\lla& not~ s, r
\end{eqnarray*}
{\em Stabilizing rules}:
\begin{eqnarray*}
q' &\lla& \neg r'\\ r' &\lla& \neg q'\\ s' &\lla& q'\\
 \neg q' &\lla& \neg s\\
  s' &\lla&
r'\\
 \neg r' &\lla& \neg s'
\end{eqnarray*}
{\em Persistence rules}:
\begin{eqnarray*}
 q'&\lla& q, not~ \neg q'\\
  s'&\lla& s, not~ \neg s'\\
   r'&\lla& r, not~ \neg r'\\
    \neg q' &\lla& not~ q, not~ q'\\
     \neg s'&\lla& not~ s, not~ s'\\
      \neg r'&\lla& not~ r, not~ r'.
\end{eqnarray*}
The answer sets are: $\{q', s', \neg r'\}$ and $\{\neg q', s',
r'\}$. Then $\nit{Core}(\Pi(r)) = \{s'\}$, but $W_{\Pi(r)} =
\emptyset$. \boxtheorem\\
\end{example}

The results obtained so far in this section apply to the repair
program $\Pi(r)$. Nevertheless, when we add an arbitrary query
program $\Pi(Q)$ to $\Pi(r)$, obtaining program $\Pi$, then it is
possible that $\nit{Core}(\Pi)$ properly extends the well-founded
interpretation of $\Pi$, even for FDs.

\begin{example} Consider $r = \{P(a,b), P(a,c)\}$,  the FD:~
$P(x,y) \vee P(x,z) \vee y = z$, and the query $Q(x): ~\exists
y~P(x,y)$. Program $\Pi$ is:

\begin{eqnarray*}
dom(a) &\leftarrow& \\ dom(b) &\leftarrow& \\dom(c) &\leftarrow&\\
P(a,b) &\leftarrow& \\ P(a,c) &\leftarrow& \\ \nit{Query}(X)
&\leftarrow& P'(X,Y)\\ P'(X,Y) &\leftarrow& P(X,Y), not~ \neg
P'(X,Y)\\
 \neg P'(X,Y) &\leftarrow& dom(X), dom(Y), not~ P(X,Y), not~
 P'(X,Y)
\end{eqnarray*}

 \begin{eqnarray*}
  \neg P'(X,Y) \vee \neg P'(X,Z) &\leftarrow& P(X,Y), P(X,Z), Y \neq
Z\\
 \neg P'(X,Y) &\leftarrow& dom(X), dom(Y), dom(Z), P'(X,Z),  Y
\neq  Z\\ \neg P'(X,Z) &\leftarrow& dom(X), dom(Y), dom(Z),
P'(X,Y), Y \neq Z.
\end{eqnarray*}

The answer sets of $\Pi$ are $S_1 = \{\it{Query}(a), P'(a,b),
P(a,b), P(a,c), \ldots\}$ and $S_2 = \{\it{Query}(a), P'(a,c),
P(a,b), P(a,c), \ldots\}$. The well-founded interpretation is
~$W_\Pi $ $= <W^+, W^-, W^u>$, with $W^+ = \{P(a,b), P(a,c),
dom(a), \ldots\}$, $W^- = \{\neg P'(a,a),$ $ \dots\}$, and
implicitly, the set of undetermined literals $W^u = \{P'(a,b),
P'(a,c),$ $ \nit{Query}(a)\}$. In particular, $\nit{Query(a)} \in
\nit{Core}(\Pi)$, but $\nit{Query}(a) \notin W^+$. \boxtheorem\\
\end{example}

We know, by complexity results presented in \cite{icdt01} for
functional dependencies that, unless $P = NP$, consistent answers
to first-order queries cannot be computed in polynomial time. In
consequence, we cannot expect to compute
 the $\nit{Core}(\Pi)$ of the program
that includes the query program by means of  the well-founded
interpretation of $\Pi$ only.

\section{Evaluating $\allrep$-queries}\label{sec:eval}

The results in section \ref{sec:bics} provide the underpinning of
a general method of evaluating $\allrep$-queries. Assume $r$ is a
database instance and the set of integrity constraints {\em IC} is
given. We show how to evaluate queries of the form $\beta\equiv
\allrep\alpha$ where $\alpha$ is a basic query. First, from
$\alpha$ we obtain a stratified logic program $\Pi(\alpha)$ (this
is a standard construction \cite{lloyd87,AbHuVi95}) in terms of
the new, primed predicates. One of the predicate symbols,
$\nit{Query}_{\alpha}$, of $\Pi(\alpha)$ is designated as the
query predicate. This is illustrated   in section
\ref{sec:queryp}. Second, determine all the answers sets
$S_1,\ldots,S_k$ of the logic program  $\Pi = \Pi(\alpha) \cup
\Pi(r)$. Third, compute the intersection~ $r_\beta=\bigcap_{1\leq
i\leq k} S_i/{\it Query}_\alpha$,~ where $S_i/{\it Query}_\alpha$
is the extension of ${\it Query}_\alpha$ in $S_i$. The set of
tuples $r_\beta$ is the set of answers to $\beta$ in $r$.

Notice that the set $U$ consisting of all the ground primed
database literals, $(\neg) p'(a)$, and all the ground non primed
database literals, $(\neg) p(a)$, form a {\em partition}
 for the program $\Pi$, because whenever a literal
in $U$ appears in a head, all the literals in the body also appear
in $U$ \cite{lifsTur94}. The set $U$ partitions the program
precisely into the two expected parts, $\Pi(r)$ and $\Pi(Q)$,
because the literals in $U$ do not appear in heads of rules in
$\Pi(Q)$ (for $\Pi(Q)$, literals in $U$ are like  extensional
literals). From \cite{lifsTur94}, we know that every answer set of
$\Pi$ can be represented as the union of an answer set of $\Pi(r)$
and an answer set of $\Pi(Q)$, where each answer set for $\Pi(r)$
acts as an extensional database for the computation of the answer
sets of $\Pi(Q)$. Program $\Pi(Q)$ is stratified, in consequence,
for each answer set for $\Pi(r)$, there will only one answer set
for $\Pi(Q)$.

To obtain query answers to general $\allrep$-queries the above
method needs to be combined with some method of evaluating
first-order queries. For example, safe-range first-order queries
\cite{AbHuVi95} can be translated to relational algebra. The same
approach can be used for $\allrep$ queries with the subqueries of
the form $\allrep\alpha$ replaced by new relation symbols. Then
when the resulting relational algebra query is evaluated and the
need arises to materialize one of the new relations, the above
method can be used to accomplish that goal.

\section{Computational Examples}\label{sec:impl}

As shown in section \ref{sec:fddbs}, our disjunctive programs with
exceptions can be transformed \cite{ks91} into extended
disjunctive  logic programs with an answer set semantics
\cite{glELPb}. Once this transformation has been performed,
obtaining program $\Pi(r)$, it is possible to  use any
implementation of extended disjunctive logic programs with answer
set semantics. In this section, we give some examples  that show
the application of the  {\em DLV} system \cite{E*98} to the
computation of database repairs and consistent query answers.

In \cite{infComp} it is shown how to compute the answer sets of a
program starting from the well-founded interpretation, that can be
efficiently computed and is contained in the intersection of the
answer sets. This is what {\em DLV} basically does, but instead of
starting from the well-founded interpretation, it starts from the
also efficiently computable set of {\em deterministic
consequences} of the program, that is still contained in the
intersection of all answer sets, and in its turn, contains the
well-founded interpretation \cite{pers}. Actually, {\em DLV} can
be explicitly asked to return the set of deterministic
consequences of the program\footnote{ By means of its option {\tt
- det}.}, and it can be also used as an approximation from below
to the intersection of all answer sets.

\subsection{Computing database repairs with {\em DLV}}

\begin{example}\label{ex:ssn} Consider a database schema ${\it Emp}({\it Name},
{\it SSN})$. Each person should have just one SSN and different
persons should have different SSNs. That is, the following
functional dependencies are expected to hold: ~${\it Name}
\rightarrow {\it SSN}$, ~${\it SSN} \rightarrow {\it Name}$. The
following is an inconsistent instance:

\begin{center}
\begin{tabular}{c|cc}
${\it Emp}$ & ${\it Name}$ & {\it SSN}\\\hhline{-|--|} &Irwin
Koper & 677-223-112\\ & Irwin Koper & 952-223-564\\ & Michael
Baneman & 334-454-991\\\hhline{~|--|}
\end{tabular}
\end{center}

\vspace{2mm}

In order to consistently query this database, we can generate the
following {\em DLV} program, where the prime predicate ${\it
Emp}^\prime$, containing the repaired extension, we had before is
now denoted by $emp\_p$

{\small
\begin{verbatim}
% domains of the database
dom_name("Irwin Koper").        dom_name("Michael Baneman").
dom_number("677-223-112").      dom_number("952-223-564").
dom_number("334-454-991").

% initial database
emp("Irwin Koper", "677-223-112").
emp("Irwin Koper", "952-223-564").
emp("Michael Baneman", "334-454-991").

% default rules
emp_p(X,Y)  :- emp(X,Y), not -emp_p(X,Y).
-emp_p(X,Y) :- dom_name(X), dom_number(Y), not emp(X,Y), not emp_p(X,Y).

% triggering rules
-emp_p(X,Y) v -emp_p(X,Z) :- emp(X,Y), emp(X,Z), Y!=Z.
-emp_p(Y,X) v -emp_p(Z,X) :- emp(Y,X), emp(Z,X), Y!=Z.

% stabilizing rules.
-emp_p(X,Y) :- emp_p(X,Z), dom_number(Y), Y!=Z.
-emp_p(Y,X) :- emp_p(Z,X), dom_name(Y), Y!=Z.
\end{verbatim}}

{\em DLV} running on this program delivers two answer sets, corresponding to
the two following repairs:

\ignore{
\begin{center}
\begin{tabular}{llcll}\hhline{|--|~|--|}
\multicolumn{2}{c}{\ssn} &\phantom{MMMMMMMM}&
\multicolumn{2}{c}{\ssn}\\\hhline{|--|~|--|}
Irwin Kuper & 677-223-112 && Irwin Kuper & 952-223-564\\
Michael Baneman & 334-454-991 &&Michael Baneman & 334-454-991\\
\hhline{|--|~|--|}
\end{tabular}
\end{center}
}

\begin{center}
\begin{tabular}{c|cc}
${\it Emp}$ & ${\it Name}$ & {\it SSN}\\\hhline{-|--|} & Irwin
Koper & 952-223-564\\ & Michael Baneman &
334-454-991\\\hhline{~|--|}
\end{tabular}
\end{center}

\vspace{2mm}
\begin{center}
\begin{tabular}{c|cc}
${\it Emp}$ & ${\it Name}$ & {\it SSN}\\\hhline{-|--|} &Irwin
Koper & 677-223-112\\ & Michael Baneman &
334-454-991\\\hhline{~|--|}
\end{tabular}
\end{center}

\vspace{2mm} In order to pose the  query ~${\it Emp}(X,Y)$?,
asking for the consistent tuples in table ${\it Employee}$, it is
necessary to add a new rule to the program:

{\small
\begin{verbatim}
query(X,Y) :- emp_p(X,Y).
\end{verbatim}}

\noindent
Now, the two answer sets of the program will contain {\tt query} literals,
namely

{\small
\begin{verbatim}
{ ..., query("Irwin Koper","952-223-564"),
                          query("Michael Baneman","334-454-991")}

{ ..., query("Irwin Koper","677-223-112"),
                          query("Michael Baneman","334-454-991")}
\end{verbatim}}

In order to obtain the consistent answers to  the query, it is
sufficient to choose all the ground {\tt query} atoms that are in
the intersection of all answer sets of the program extended by the
query rule. In this case, we obtain as only consistent answer the
tuple: \verb+X="Michael Baneman"+, \verb+Y="334-454-991"+. Here we
had a non-quantified conjunctive query. In other cases, the {\tt
query} predicate will be defined by a more complex program
$\Pi(Q)$.  \hfill $\Box$
\end{example}

\section{An Alternative Semantics} \label{sec:belief}

As discussed in \cite{pods99}, our notion of database repair
coincides with that of  revision model obtained with the
``possible model approach" introduced by Winslett in
\cite{winslett88} (see also \cite{winslett94}) in the context of
belief update, when the database instance (a model) is updated by
the ICs, generating a new set of models, in this case, the
database repairs. In consequence, we have shown in section
\ref{sec:fddbs} that our repair program $\Pi(r)$ ~ (cf. theorem
\ref{tw})~ has as its answer sets the Winslett's revision models
of $r$ wrt $IC$.

\subsection{Cardinality based repairs and weak constraints}\label{sec:dalal's}

Winslett's revision models are based on minimal {\em set} of changes.
In \cite{dalal}, Dalal presents an alternative notion of revision model based on minimal
{\em number} of changes.

\begin{definition}
Given a database instance $r$, an instance $r^\prime$ is a {\em
Dalal repair} of $r$ wrt to ${\it IC}$ iff $r^\prime\models {\it
IC}$ and $|\Delta(r,r^\prime)|$ is a minimal element of
$\{|\Delta(r,r^*)|\mid r^*\models IC\}$. \boxtheorem\\
\end{definition}

We could give a definition of {\em Dalal consistent answer}
exactly in the terms of definition \ref{def:basic}, but replacing
``repair" by ``Dalal repair".

It is possible to specify Dalal repairs  using  the same repair
programs we had in section \ref{sec:bics}, but with the
persistence defaults replaced by {\em weak constraints}
\cite{weak}. The latter will not be a sort of weak version of the
original, database ICs. Rather they will be new constraints
imposed on the answer sets of the repair program, actually on
$\Pi_\Delta(r)$, the part of the repair program of section
\ref{sec:bics} that is responsible for the changes.

As described in \cite{weak}, weak constraints are written in the
form ~$\Leftarrow L_1, \ldots, L_n$, where the $L_i$'s are
literals containing strong or weak negation. They are added to an
extended disjunctive program. Their semantics is such, that, when
violated in a model of the program, they do not necessarily
``kill" the model. The models of the program that minimize the
{\em number} of violated ground instantiations of the given weak
constraints are kept.

In order to capture the Dalal repairs we need a very simple form
of weak constraint. The program $\Pi^{D}(r)$ that specifies the
Dalal repairs of a database instance $r$ wrt a set of BICs
consists of program $\Pi_{\Delta}(r)$ of section \ref{sec:fddbs}
(rules 1. -- 3.) plus

\begin{itemize}
\item [4".] For every database predicate $p$, the weak
constraints
\begin{eqnarray}
 &\Leftarrow& ~p^\prime(\bar{x}), ~{\it
not}~ p(\bar{X}), \label{eq:wics}\\ &\Leftarrow& ~\neg
p^\prime(\bar{x}), ~ p(\bar{X}). \nonumber
\end{eqnarray}
\end{itemize}

These constraints say that the contents of the original database
and of each repair are expected to coincide. Since they are weak
constraints, they allow violations, but only a minimum {\emph
number} of tuples that belong to the repair and not to the
original instance, or the other way around, will be accepted.

The results for the change program $\Pi_\Delta(r)$ still hold
here. The program obtained by the combination of the change
program with the weak constraints 4". in (\ref{eq:wics}) will have
answers sets that correspond to repairs that are minimal under set
inclusion and under number of changes, i.e. Dalal repairs only.
So, for BICs and finite domain databases we have

\begin{theorem}\label{td}
Given a (finite domain) database instance $r$ and a set of BICs
~${\it IC}$:
\begin{enumerate}
\item For every Dalal repair $r^\prime$ of $r$ wrt ${\it IC}$, there exists an answer
set ~$S$~ of $\Pi^D(r)$ such that $I(S)=r^\prime$.
\item For every answer set $S$ of $\Pi^D(r)$, there exists a Dalal repair
$r^\prime$ of $r$ wrt ${\it IC}$ such that $I(S)=r^\prime$.
\boxtheorem
\end{enumerate}
\end{theorem}

As with Winslett's repairs, the theorem still holds for infinite
domain databases when the BICs are domain independent.

\begin{example} Let $D=\{a\}$, $r=\{p(a)\}$ and ${\it IC} =\{\neg p(x) \vee
q(x), \neg q(x) \vee r(x)\}$. In this case, $\Pi^D(r)$ is
\begin{eqnarray*}
{\it dom}(a) &\lla&\\ p(a) &\lla&\\ \neg p^\prime(X) \vee
q^\prime(X) ~&\lla& ~p(X), ~{\it not}~ q(X)\\ q^\prime(X) ~&\lla&~
p^\prime(X)\\ \neg p^\prime(X) ~&\lla&~ \neg q^\prime(X)\\ \neg
q^\prime(X) \vee r^\prime(X) ~&\lla&~ q(X), ~{\it not}~ r(X)\\
r^\prime(X) ~&\lla&~ q^\prime(X)\\ \neg q^\prime(X) &\lla& \neg
r^\prime(X)\\ &\Leftarrow&~ p^\prime(X), ~{\it not}~ p(X)\\
&\Leftarrow&~ \neg p^\prime(X), p(X)\\ &\Leftarrow&~ q^\prime(X),
~{\it not}~ q(X)\\ &\Leftarrow&~ \neg q^\prime(X), q(X)\\
&\Leftarrow&~ r^\prime(X), ~{\it not}~ r(X)\\ &\Leftarrow&~ \neg
r^\prime(X), r(X).
\end{eqnarray*}

Weak constraints are implemented in {\em DLV}\footnote{They are
and  are specified by ~$:\sim ~{\it Conj.}$, where ${\it Conj}$ is
a conjunction of (possibly negated) literals. See {\em DLV}'s user
manual in ~http://www.dbai.tuwien.ac.at/proj/dlv/man.}. Running
{\em DLV} on this program we obtain the answer set $\{{\it
dom}(a), p(a),$ $ \neg p^\prime(a)\}$, corresponding to the empty
database repair, but not
 the other Winslett's repair ~$\{p(a), q(a), r(a)\}$, whose set of
 changes wrt $r$ has two elements, whereas the first repair
 differs from $r$ by one change only. \boxtheorem\\
\end{example}

Notice that, by construction of $\Pi^D(r)$, in most cases the
primed predicates in the
 answer sets of the program  will not contain all the information (we replaced the persistence defaults
 4. by the weaks constraints 4".). In consequence, we will have to interpret the
result, and the answers to queries will be obtained by using
negation as failure, as we shown in the following table:\\

\begin{center}
\begin{tabular}{|c|c|}\hhline{|--|}
original query    & query in the program\\\hhline{|==|}
$p(\bar{x})$      & $\tt query(\bar{X}) \di p^\prime(\bar{X}).$\\
                  & $\tt query(\bar{X}) \di p(\bar{X}) \dc \dnf \neg
                  p^\prime(\bar{X}).$\\\hhline{|--|}
$\neg p(\bar{x})$ & $\tt query(\bar{X}) \di \neg
p^\prime(\bar{X}).$\\
                  & $\tt query(\bar{X}) \di {\it dom}(\bar{X}) \dc \dnf
p(\bar{X}) \dc \dnf p^\prime(\bar{X}).$\\\hhline{|--|}
\end{tabular}
\end{center}

\section{Extensions}\label{sec:remarks}

In this section we will show how the specifications of database
repairs given for binary integrity constraints can be extended to
referential integrity constraints and to universal constrains with
more than two database literals in the standard format. We will
only consider the case of minimal repairs under set inclusion.

\subsection{Referential integrity constraints} \label{sec:ref}

By appropriate representation of existential quantifiers as
program rules it is possible to apply the methodology for
universal constraints presented in the previous sections to handle
referential integrity constraints (RICs).

Consider the {\em RIC}: ~$\forall \bar{x}~(P(\bar{x}) \rightarrow
\exists \bar{y} ~R(\bar{x},\bar{y}))$, and the inconsistent
database instance $r = \{P(\bar{a}), P(\bar{b}),
R(\bar{b},\bar{a})\}$. For things to work properly, it is
necessary to assume that there is an underlying database domain
$D$. The repair program has the persistence default rules
$$P'(\bar{X}) \leftarrow P(\bar{X}); ~~~\neg P'(\bar{X})
\leftarrow \it{not}~ P(\bar{X});$$ $$R'(\bar{X},\bar{Y})
\leftarrow R(\bar{X},\bar{Y});~~~ \neg R'(\bar{X},\bar{Y})
\leftarrow \it{not}~ R(\bar{X},\bar{Y}).$$ In addition, it has the
stabilizing exceptions
\begin{eqnarray}
\neg P'(\bar{X}) &\leftarrow& {\it not}~ {\it aux}'(\bar{X}), \neg
R'(\bar{X},{\it null}), \label{eq:null}\\ R'(\bar{X},{\it null})
&\leftarrow& P'(\bar{X}), {\it not}~ {\it aux}'(\bar{X});
\label{eq:null2}
\end{eqnarray}
with $${\it aux}'(\bar{X}) ~\leftarrow~ R'(\bar{X},\bar{Y});$$ and
the triggering exception \begin{equation}\neg P'(\bar{X}) \vee
R'(\bar{X},{\it null}) \leftarrow P(\bar{X}), {\it not}~ {\it
aux}(\bar{X}), \label{eq:tri}\end{equation}
 with ~${\it aux}(\bar{X})
~\leftarrow~ R(\bar{X},\bar{Y})$.

The variables in this program range over $D$, that is, they do not
take the value ${\it null}$. This is the reason for the last
literal in clause (\ref{eq:null}). The last literal in clause
(\ref{eq:null2}) is necessary to insert a null value only when it
is needed; this clause relies on the fact that variables range
over $D$ only. Instantiating  variables on $D$ only\footnote{A
simple way to enforce this at the object level is to introduce the
predicate $D$ in the clauses, to force variables to take values in
$D$ only, excluding the null value.}, the only two answer sets are
the expected ones, namely delete $P(\bar{a})$ or insert
$R(\bar{a}, {\it null})$.

It would be natural  to include here the functional dependency
$\bar{X} \rightarrow \bar{Y}$ on $R$, expressing that $\bar{X}$ is
a primary key in $R$ and a foreign key in $P$. This can be done
without problems, actually the two constraints would not interact,
that is, reparing one of them will not cause violations of the
other one.

Finally, it should be clear how to modify the specification above
if the only admissible changes are elimination of tuples, but not
introduction of null values. For example, the triggering exception
(\ref{eq:tri}) would have to be changed into ~$\neg P'(\bar{X})
\leftarrow P(\bar{X}), {\it not}~ {\it aux}(\bar{X})$.

\subsection{Referential ICs and strong constraints}\label{sec:strong}

It has been possible to use {\em DLV} to impose some preferences
on the repairs via an appropriate representation of constraints,
obtaining, for example for RICs, introduction of null values, a
cascade policy, ...

\begin{example} (example \ref{ex:ssn} cont'd)  Consider the same schema and {\em FD}s as before,
but now extended with a unary table ${\it Person(Name)}$. Now, we
have the following instance

\begin{center}
\begin{tabular}{c|cc}
${\it Emp}$ & ${\it Name}$ & ${\it SSN}$\\\hhline{|---|} & Irwin
Koper & 677-223-112\\ & Irwin Koper & 952-223-564\\ & Michael
Baneman & 952-223-564\\\hhline{~|--|}
\end{tabular}
\end{center}

\vspace{2mm}
\begin{center}
\begin{tabular}{c|c}
${\it Person}$ & ${\it Name}$\\\hhline{|--|} & Irwin Koper\\ &
Michael Baneman\\\hhline{~|-|}
\end{tabular}
\end{center}

\vspace{2mm} The {\em DLV} repair program, without considering any
change on table {\em Person}, is as in example \ref{ex:ssn}, but
with:

{\small
\begin{verbatim}
dom_number("677-223-112").         dom_number("952-223-564").
% initial database
emp("Irwin Koper", "677-223-112"). emp("Irwin Koper",
"952-223-564"). emp("Michael Baneman", "952-223-564").
\end{verbatim}}

\noindent If {\em DLV} is run with this program as input, we
obtain the answer sets:

{\small
\begin{verbatim}
{ ..., emp_p("Irwin Koper","677-223-112"),
      -emp_p("Irwin Koper","952-223-564"),
       emp_p("Michael Baneman","952-223-564"),
      -emp_p("Michael Baneman","677-223-112")}

{ ...,-emp_p("Irwin Koper","677-223-112"),
       emp_p("Irwin Koper","952-223-564"),
      -emp_p("Michael Baneman","952-223-564"),
      -emp_p("Michael Baneman","677-223-112")}
\end{verbatim}}

\noindent corresponding to the database repairs:

\begin{center}
\begin{tabular}{c|cc}

${\it Emp}$ & ${\it Name}$ & ${\it SSN}$\\\hhline{|---|} & Irwin
Koper & 677-223-112\\ & Michael Baneman &
952-223-564\\\hhline{~|--|}
\end{tabular}
\end{center}

\vspace{2mm}
\begin{center}
\begin{tabular}{c|cc}
${\it Emp}$ & ${\it Name}$ & ${\it SSN}$\\\hhline{|---|} &
~~~~Irwin Koper~~~~ & 952-223-564\\\hhline{~|--|}
\end{tabular}
\end{center}

\vspace{2mm} In the second repair Michael Baneman does not have a
SSN, in consequence it is not a consistent answer that he has a
SSN. Actually, it is possible to ask with {\em DLV} for those
persons who  have a SSN by computing the answer sets of  the
program extended by the query rule: ~\verb+query(X):- emp_p(X,Y).+
Two answer sets are obtained:

{\small
\begin{verbatim}
{ ..., query("Irwin Koper"), query("Michael Baneman")}, { ...,
query("Irwin Koper")}
\end{verbatim}}

From them, we can say that only Irwin Koper has a SSN (in all
repairs). If we want every person to have a SSN, then we may
enforce the RIC,  ~$\forall ~x({\it Person}(x) ~\rightarrow$
$\exists ~y {\it Emp}(x,y))$, stating that every person must have
a SSN.

In section \ref{sec:ref}, we repaired the database introducing the
RIC as a part of the program, producing either the introduction of
null values or cascading deletions. We may  not want any of these
options (we do not want null values in the key {\it SSN}) or we do
not want to delete any employees (in this case, M. Baneman from
{\it Person}). An alternative is to use {\em DLV}'s possibility of
specifying {\em strong constraints}, that have the effect of
pruning those answer sets that do not satisfy them. This can be
                   done in {\em DLV} by introducing the denial
                   \verb+:- dom_name(X), not has_ssn(X).+,
                   with ~ \verb+has_ssn(X) :- emp_p(X,Y).+. The
answer sets of the original program that do not satisfy the ICs
are filtered out.

Now, only one repair is obtained:

{\small
\begin{verbatim}
{ ..., emp_p("Irwin Koper","677-223-112"),
      -emp_p("Irwin Koper","952-223-564"),
       emp_p("Michael Baneman","952-223-564"),
      -emp_p("Michael Baneman","677-223-112"),
       has_ssn("Irwin Koper"),  has_ssn("Michael Baneman"),
       query("Irwin Koper"),  query("Michael Baneman")}
\end{verbatim}}

In it, every person has a SSN (according to the {\tt has\_ssa}
predicate). As expected, the answers to the original query are
\verb+X="Irwin Koper"+ and \verb+X="Michael+ \verb+Baneman"+.
\end{example}

\subsection{Extensions to other universal ICs}

In order to handle universal ICs with a larger number of database
literals, we can use the same program $\Pi(r)$, but with
$\Pi_\Delta(r)$, as introduced in definition \ref{def:changeprogs}
for BICs, now  without the condition $n + m \leq 2$ in
(\ref{eq:genBics}). Now $m$ and $n$ can be arbitrary. This will
produce properly disjunctive stabilizing rules, by passing in
turns just one of the disjuncts to the body in it complementary
form. Furthermore, we need to extend $\Pi_\Delta(r)$ with more
disjunctive stabilizing rules. In essence, as indicated in
\cite{fqas2k}, we need to consider all possible subsets of the
database literals appearing  in the standard format
(\ref{eq:genBics}) and put them in a disjunction in the head,
passing to the body the remaining literals and the formula
$\varphi$. We will show this by means of some examples.

\ignore{
\begin{example} \label{ex:counter}
Let us take ~$D=\{a\}$, $r=\{p(a), r(a)\}$ and ${\it IC}$
consisting of:~ $\neg p(x) \vee \neg q(x) \vee r(x)$ and $\neg r(x) \vee \neg x=a$.
In this case, $\Pi_\Delta(r)$ is
\begin{itemize}
\item []${\it dom}(a), ~p(a), r(a)$.

\item []$\neg p^\prime(X) \vee \neg q^\prime(X) \vee r^\prime(X) \lla p(X), q(X), {\it
not}~
r(X)$\\
$\neg p^\prime(X) \vee \neg q^\prime(X) \lla \neg r^\prime(X)$\\
$\neg p^\prime(X) \vee r^\prime(X) \lla q^\prime(X)$\\
$\neg q^\prime(X) \vee r^\prime(X) \lla p^\prime(X)$

\item []$\neg r^\prime(X) \lla r(X), X=a$\\
$\neg r^\prime(X) \lla {\it dom}(X), X=a$.
\end{itemize}

In this case, there
is only one repair of $r$, namely ~$r^\prime=\{p(a)\}$. But,
the program has as answer sets~
$\{{\it dom}(a), p(a), r(a), \neg r^\prime(a), \neg q^\prime(a)\}$
and $\{{\it dom}(a), p(a),$ $ r(a), \neg r^\prime(a), \neg
p^\prime(a)\}$.
The first of them represents $r^\prime$, but the second one represents the
empty database instance, which satisfies ${\it IC}$, but is not a repair of $r$.
\boxtheorem
\end{example}
}

\begin{example}\label{ex:ternary}
Consider the DB instance~ $r=\{P(a),Q(a),R(a)\}$~ and the
following set of ternary integrity constraints~~$IC = \{\neg P(x)
\vee \neg Q(x) \vee R(x), \neg P(x) \vee \neg Q(x) \vee \neg R(x),
\neg P(x) \vee Q(x) \vee \neg R(x), P(x) \vee \neg Q(x) \vee \neg
R(x), \neg P(x) \vee Q(x) \vee R(x), P(x) \vee \neg Q(x) \vee
R(x), P(x) \vee Q(x) \vee \neg R(x)\}$.~ The repair program
contains the usual persistence defaults for $P, Q, R$, plus
triggering exceptions, e.g. for the first IC in $IC$:
\begin{eqnarray*}
\neg P'(x) \vee   \neg Q'(x) \vee   R'(x) &\leftarrow&  P(x),
Q(x), not~ R(x), \label{eq:te1}
\end{eqnarray*}
and the stabilizing rules directly obtained from definition
\ref{def:changeprogs}, e.g. for the first IC
\begin{eqnarray}
\neg P'(x) \vee \neg Q'(x) &\leftarrow& \neg R'(x),
\label{stadis}\\
 ~\neg P'(x)
\vee R'(x) &\leftarrow& Q'(x), \nonumber\\
 ~\neg Q'(x) \vee R'(x) &\leftarrow&
P'(x). \nonumber
\end{eqnarray}
We add to the program the other possible combinations, e.g. for
the first IC:
\begin{eqnarray}
\neg  P'(x)  &\leftarrow&  Q'(x), \neg R'(x),\label{stanodis}\\
R'(x) &\leftarrow&
 P'(x), Q'(x), \nonumber \\
 \neg Q'(x) &\leftarrow& P'(x), \neg R'(x).\nonumber
\end{eqnarray}
In this case we obtain as answer set  the only repair, namely the
empty instance, represented by $\{P(a), Q(a), R(a), \neg P'(a),
\neg Q'(a), \neg R'(a)\}$. It is easy to verify that  without
using the disjunctive stabilizing rules (\ref{stadis}), but with
the rules  (\ref{stanodis}) as the only stabilizing rules, the
empty repair cannot be obtained. \boxtheorem
\end{example}

\begin{example} Let $D=\{a,b,c\}$, $r=\{p(a,b),p(b,c)\}$ and
${\it IC}$  the transitivity constraint~ $\neg p(x,y) \vee \neg
p(y,z) \vee p(x,z)$. Here, the only repairs are $r_1=\{p(a,b)\}$,
$r_2=\{p(b,c)\}$ and $r_3=\{p(a,b),p(b,c),p(a,c)\}$. The  repair
program $\Pi(r)$ contains the facts~ $dom(a), dom(b),$ $ dom(c),
p(a,b), p(b,c)$ ~plus the rules
\begin{eqnarray*}
  p'(X,Y) &\leftarrow& p(X,Y), not~ \neg p'(X,Y)\\
\neg p'(X,Y) &\leftarrow& dom(X), dom(Y), not~ p(X,Y), not~
p'(X,Y)\\ \neg p'(X,Y) \vee \neg p'(Y,Z) \vee p'(X,Z) &\leftarrow&
p(X,Y), p(Y,Z), not~ p(X,Z) \nonumber\\
\end{eqnarray*}

\begin{eqnarray}
\neg p'(Y,Z) \vee p'(X,Z)
&\leftarrow& dom(Z), p'(X,Y) \label{eq:dist}\\ \neg p'(X,Y) \vee
p'(X,Z) &\leftarrow& dom(X), p'(Y,Z) \nonumber\\
 \neg p'(X,Y) \vee \neg p'(Y,Z) &\leftarrow& dom(Y),
\neg p'(X,Z) \nonumber\\ \neg p'(Y,Z) &\leftarrow& dom(Z),
p'(X,Y), \neg p'(X,Z) \nonumber\\
 p'(X,Z) &\leftarrow& p'(X,Y), p'(Y,Z) \nonumber\\
  \neg p(X,Y) &\leftarrow& dom(X), p'(Y,Z), \neg p'(X,Z).
  \nonumber
\end{eqnarray}
With this program we will get as answer sets the three repairs,
but also six other models corresponding to non-minimal repairs,
e.g. ~~$\{dom(a),$ $dom(b),$ $dom(c),$ $p(a,b),$ $p(b,c),$
$p'(a,b),$ $p'(b,c),$ $p'(a,c),$ $p'(b,b),$ $p'(c,b),$ $p'(c,c),$
$p'(a,a),$ $p'(b,a),$ $p'(c,a)\}$, meaning that we are repairing
the database by inserting the tuple $(a,c)$ into $p$, but also the
tuples $(b,b), (c,b)$, etc. The reason is that a rule like
(\ref{eq:dist}) for example is inserting tuples $(x,z)$, for
essentially any $z$, independently of the condition in the body.
In order to solve this problem, one can add extra conditions in
the body of (\ref{eq:dist}) that relate  the possible values for
$z$ with the values  in the original database. In consequence,
that rule should be replaced by ~$\neg p'(Y,Z) \vee p'(X,Z)
\leftarrow p(Y,Z), not~ p(X,Z), p'(X,Y)$. The domain predicate,
introduced in (\ref{eq:dist}) to make the rule safe, is no longer
needed. Similar changes have to be performed in the other
stabilizing rules. \boxtheorem
\end{example}

\section{Conclusions} \label{sec:concl}

There are several interesting open issues related to computational
implementation of the methodology we have presented.

The existing implementations  of stable models semantics are based
on grounding the rules, what, in database applications, may lead
to huge ground programs. In addition, those implementations are
geared to computing stable models, possibly not all of them, and
answering ground queries. At the same time, in database
applications the possibility of posing and answering open queries
(with variables) is much more natural. In addition, consistent
query answering requires, at least implicitly, having all stable
models.

It would be useful to implement a consistent query answering
system based on the interaction of our repairs logic programs with
relational DBMS.  For this purpose, some functionalities and
front-ends included in {\em DLV}'s architecture \cite{dlvSys}
could be  used.

Another interesting issue has to do with the possibility of having
the consistent query answering mechanism guided by the query, so
that irrelevant computations are avoided.

There, are other open problems that could be considered: (a)
Analyzing conditions under which simpler programs can be obtained.
(b) A more detailed treatment of  referential ICs (and other
existential ICs). (c) Identification of other classes of ICs for
which the well-founded interpretation and the intersection of all
database repairs coincide. (d) Preferences for certain kinds of
repair actions. In principle they could be captured by choosing
the right disjuncts in the triggering rules.

With respect to related work, the closest approach to ours is
presented in \cite{greco3} (see also \cite{greco1,greco2}). There
disjunctive programs are used to specify the sets of changes under
set inclusion that lead to database repairs in the sense of
\cite{pods99}. They present a compact schema for generating repair
programs for universal integrity constraints. The application of
such a schema leads to programs that involve essentially all
possible disjunctions of database literals in the heads. They
concentrate mainly on producing the set of changes, rather than
the repaired databases explicitly. In particular, they do no have
persistence rules in the program. In consequence, the program
cannot be used directly to obtain consistent answers. They also
introduce ``repair constraints" to specify preferences for certain
kinds of repairs.

Another approach to database repairs based on logic programming
semantics consists of the {\em revision programs} \cite{mt}. The
rules in those programs explicitly declare how to enforce the
satisfaction of an integrity constraint, rather than explicitly
stating the ICs,  e.g. $$\nit{in}(a) \leftarrow \nit{in}(a_1),
\ldots, \nit{in}(a_k), \nit{out}(b_1), \ldots, \nit{out}(b_m)$$
has the intended procedural meaning of inserting the database atom
$a$ whenever $a_1, \ldots, a_k$ are in the database, but not $b_1,
\ldots, b_m$. They also give a declarative, stable model semantics
to revision programs. Preferences for certain kinds of repair
actions can be captured by declaring the corresponding rules in
program and omitting rules that could lead to other forms of
repairs. Revision programs could be used, as the programs in
\cite{greco3}, to obtain consistent answers, but not directly,
because they give an account of the changes only.

\vspace{4mm}\noindent {\bf Acknowledgments:}~ Work supported by
FONDECYT Grant 1000593, NSF Grant INT-9901877/CONICYT Grant
1998-02-083, NSF Grant IIS-0119186, Carleton University Start-Up
Grant 9364-01. We are grateful to Francisco Orchard for
informative paper presentations, discussions, and experiments with
DLV. We are grateful to Nicola Leone for kindly answering all our
questions about {\em DLV}.

\appendix
\section*{Appendix: Proofs}

\Proof{Proposition \ref{l1}} Consider an arbitrary element in
$\nit{IC}$:
\begin{eqnarray*}
&&\bigvee_{i=1}^n p_i(\bar{x}_i)\vee\bigvee_{i=1}^m \neg
q_i(\bar{y}_i)\vee \varphi.
\end{eqnarray*}
 We have to prove
that $I(S)$ satisfies any instantiation of this formula, that is
\begin{eqnarray}\label{e1}
&&I(S)\models\bigvee_{i=1}^n p_i(\bar{a}_i)\vee\bigvee_{i=1}^m
\neg q_i(\bar{b}_i)\vee \varphi
\end{eqnarray}
We need to consider two cases.
\begin{enumerate}
\item[(I)] If $r$ does not satisfy this ground constraint, then
$S$ satisfies the body of the ground triggering rule:
$$\bigvee_{i=1}^n p_i^\prime(\bar{a}_i) \dor \bigvee_{i=1}^m \dln
q_i^\prime(\bar{b}_i) \di dom(\bar{a}_1,\ldots,\bar{a}_n) \dc
\bigwedge_{i=1}^n \dnf p_i(\bar{a}_i) \dc \bigwedge_{i=1}^m
q_i(\bar{b}_i) \dc \dnf \varphi.$$ Thus, there exists
$p_i^\prime(\bar{a}_i)\in S$ or $\dln q_j^\prime(\bar{b}_j)\in S$.
But, if $p_i^\prime(\bar{a}_i)\in S$, then $I(S)\models
p_i(\bar{a}_i)$, and therefore (\ref{e1}) holds. If $\dln
q_j^\prime(\bar{b}_j)\in S$, then $ q_j^\prime(\bar{b}_j)\not\in
S$. Thus $I(S)\models\neg q_j(\bar{b}_j)$, and therefore
(\ref{e1}) holds.

\item[(II)] If $r$ satisfies the ground constraint, then $r$
could satisfy $\varphi$, and in this case $I(S)\models\varphi$.
Otherwise,  $r$  satisfies some $p_j(\bar{a}_j)$ or some $\neg
q_j(\bar{b}_j)$.

Assume that (\ref{e1}) is not true. In this case,
$I(S)\not\models\varphi$, and therefore $r$ must satisfy some
$p_j(\bar{a}_j)$ or some $\neg q_j(\bar{b}_j)$.  If $r$ satisfies
$p_j(\bar{a}_j)$, then $ p_j(\bar{a}_j)\in S$. But,
$I(S)\not\models p_j(\bar{a}_j)$, since (\ref{e1}) does not holds,
and therefore $\dln p_j^\prime(\bar{a}_j)\in S$, by definition of
$I(S)$. But, in this case $S$ satisfies the body of the ground
stabilizing rule:
\begin{multline*}
\bigvee_{i=1}^{j-1} p_i^\prime(\bar{a}_i) \dor \bigvee_{i=j+1}^n
p_i^\prime(\bar{a}_i) \dor \bigvee_{i=1}^m \dln
q_i^\prime(\bar{b}_i) \di\\ dom(\bar{a}_1,\ldots,\bar{b}_m) \dc
\dln p_j^\prime(\bar{a}_j) \dc \dnf \varphi.
\end{multline*}
Thus, by using an argument analogous to the argument given in (I),
we  conclude that (\ref{e1}) holds, a contradiction.

The case in which  $r$ satisfies $\neg q_j(\bar{b}_j)$ can be
handled in a similar way. \boxtheorem \ignore{ , then $
q_j(\bar{b}_j)\not\in S$. But, $I(S)\models q_j(\bar{b}_j)$, since
(\ref{e1}) does not holds, and therefore $
q_j^\prime(\bar{b}_j)\in S$, by definition of $I(S)$. But, in this
case $S$ satisfies the body of the ground stabilizing rule:
\begin{multline*}
 \bigvee_{i=1}^n p_i^\prime(\bar{a}_i) \dor \bigvee_{i=1}^{j-1} \dln
q_i^\prime(\bar{b}_i) \dor \bigvee_{i=j+1}^m \dln
q_i^\prime(\bar{b}_i) \di\\
dom(\bar{a}_1,\ldots,\bar{a}_n,\bar{b}_1,\ldots,\bar{b}_{j-1},\bar{b}_{j+1},\ldots,\bar{b}_m)
\dc q_j^\prime(\bar{b}_j) \dc \dnf \varphi.
\end{multline*}
Thus, by using again an argument analogous to the one in (I), we
 conclude that (\ref{e1}) holds, a contradiction.}
\end{enumerate}

\Proof{Proposition \ref{l2}} In order to prove that $ S(r,r')$
satisfies $\Pi_\Delta(r)$, we need  to consider only the four
different kinds of ground stabilizing rules (the satisfaction of
the other rules follows from the fact that $r'$ satisfies $IC$).

If $ S(r,r')$ satisfies the body of the rule $q^\prime(\bar{b})
\di p^\prime(\bar{a}) \dc {\it not}\ \varphi$, then $r'$ must
satisfy $p(\bar{a})$ and $\neg\varphi$. But $r'\models q(\bar{b})
\vee \neg p(\bar{a}) \vee \varphi$, because $\forall \bar{x}
\forall \bar{y} (q(\bar{x}) \vee \neg p(\bar{y}) \vee \varphi) \in
\nit{IC}$, and therefore, $r'\models q(\bar{b})$. Thus,
$q^\prime(\bar{b})\in S(r,r')$.

In the same way, it is possible to prove that $S(r, r')$ satisfies
all the rules of the form:
\begin{eqnarray*}
\dln q^\prime(\bar{b}) &\di& p^\prime(\bar{a}) \dc \dnf \varphi,\\
q^\prime(\bar{b}) &\di& \dln p^\prime(\bar{a}) \dc \dnf \varphi,\\
\dln q^\prime(\bar{b}) &\di& \dln p^\prime(\bar{a}) \dc \dnf
\varphi.
\end{eqnarray*}
\boxtheorem\\

\Proof{Proposition \ref{exstable}} From the previous proposition,
we know that the change program has models. Now, that program can
be partitioned into two programs, the first one containing the
stabilizing rules and modified versions of the triggering rules,
where the literals of the form $not~ p$ in the bodies are replaced
by $p^\star$. The other one contains the domain and database facts
plus the new rules $p^\star(\bar{X}) \leftarrow not ~p(\bar{X})$.
By a result in \cite{lifsTur94}, the partitioned program has as
answer sets the unions of the answer sets of the second program
and the answer sets of the first one, where the atoms $p^\star$
are treated as extensional database predicates. The second program
is stratified and has one answer set. The first one does not
contain weak negation, it is a positive program in that sense, and
its minimal models coincide with its answer sets. \boxtheorem\\

\Proof{proposition \ref{aux2}} Notice that the two sets added to
$S_M$ on the right-hand side are expected to give an account of
the persistence rules that are not included in $\Pi_\Delta(r)$.

Let $S'_M$ be the set added to $S_M$:
\begin{eqnarray*}
&&\{ p^\prime(\bar{a}) \mid  p(\bar{a})\in S_M \text{ and }  \dln
p^\prime(\bar{a})\not\in S_M\}\cup \{ \dln p^\prime(\bar{a}) \mid
p(\bar{a})\not\in S_M \text{ and }  p^\prime(\bar{a})\not\in
S_M\}.
\end{eqnarray*}
It is easy to verify that $^S\Pi(r) =~ ^{S_M}\Pi_\Delta(r)$. Then,
since $S_M$ is an answer set  of $\Pi_\Delta(r)$, in order to
prove that $S$ is an answer set
 of $\Pi(r)$, it suffices to prove (I) and (II) below.
\begin{enumerate}
\item[(I)] $S'_M \subseteq \cap\alpha(^S\Pi(r))$.

Let $l(\bar{a})$ be an element of $S'_M$. If $l(\bar{a}) =
p'(\bar{a})$, then $ p(\bar{a})\in S_M$ and $ \dln
p^\prime(\bar{a})\not\in S_M$, and, therefore, $p(\bar{a})$ and $
p^\prime(\bar{a}) \di p(\bar{a})$ are rules in $^S\Pi(r)$. Thus, $
p^\prime(\bar{a})$ is in $\cap\alpha(^S\Pi(r))$. If $l(\bar{a}) =
\neg p'(\bar{a})$, then $ p(\bar{a})\not\in S_M$ and $
p^\prime(\bar{a})\not\in S_M$, and, therefore, $\dln
p^\prime(\bar{a}) \di dom(\bar{a})$ is a reduced ground
persistence rule in $^S\Pi(r)$. Thus, $ \dln p^\prime(\bar{a})$ is
in $\cap\alpha(^S\Pi(r))$.

\item[(II)] From $S'_M$ is not possible to deduce an element that is
not included in $S$.

Assume that $q^\prime(\bar{Y}) \di dom(\bar{Y}) \dc
p^\prime(\bar{X}) \dc \dnf \varphi$ is a rule in $\Pi_\Delta(r)$,
and $ q^\prime(\bar{b}) \di$ $ dom(\bar{b}) \dc p^\prime(\bar{a})$
is a rule in $^S\Pi(r)$. If $ p^\prime(\bar{a})\in S'_M$, we need
to show that $ q^\prime(\bar{b})\in S$. By contradiction, suppose
that $ q^\prime(\bar{b})\not\in S$. Then $
q^\prime(\bar{b})\not\in S_M$ and $q^\prime(\bar{b}) \not\in
S'_M$. Therefore, $ q(\bar{b})\not\in S_M$ or $ \dln
q^\prime(\bar{b})\in S_M$. If $ q(\bar{b})$ is not in $ S_M$, then
$ S_M$ satisfies the body of the rule
 $$q^\prime(\bar{b}) \dor \dln
p^\prime(\bar{a}) \di dom(\bar{b}) \dc p(\bar{a}) \dc \dnf
q(\bar{b}) \dc \dnf \varphi,$$ because  $p'(\bar{a}) \in S'_M$. In
consequence, $p(\bar{a}) \in S_M$. But, this implies that $
q^\prime(\bar{b})\in S_M$, a contradiction, or $ \dln
p^\prime(\bar{a})\in S_M$, also a contradiction (since
$p^\prime(\bar{a})\in S'_M$). Otherwise, if $ \dln
q^\prime(\bar{b})\in S_M$, then by using the rule $ \dln
p^\prime(\bar{a}) \di dom(\bar{a}) \dc \dln q^\prime(\bar{b})$, we
can conclude that $ \dln p^\prime(\bar{a})$ is in $ S_M$, a
contradiction.

Analogously, it is possible to prove the same property for any of
the rules:
\begin{eqnarray*}
\dln q^\prime(\bar{Y}) &\di& dom(\bar{Y}) \dc p^\prime(\bar{X})
\dc \dnf \varphi,\\ q^\prime(\bar{Y}) &\di& dom(\bar{Y}) \dc \dln
p^\prime(\bar{X}) \dc \dnf \varphi,\\ \dln q^\prime(\bar{Y}) &\di&
dom(\bar{Y}) \dc \dln p^\prime(\bar{X}) \dc \dnf \varphi.
\hspace{4.8cm} \Box\\
\end{eqnarray*}
\end{enumerate}

\Proof{Lemma \ref{l3}}  Let $S$ be a answer set of $\Pi_\Delta(r)$
such that $S$ is a subset of $ S(r,r')$. First, we will prove that
$\Delta(r,I(S))\subseteq\Delta(r,r')$.

If $p(\bar{a})\in\Delta(r,I(S))$, then one of the following cases
holds.
\begin{enumerate}
\item[(I)] $r\models p(\bar{a})$ and $I(S)\not\models p(\bar{a})$. In
this case, $ p(\bar{a})\in S$ and $ p^\prime(\bar{a})\not\in S$.
Thus, by definition of $I(S)$ we conclude that $ \dln
p^\prime(\bar{a})\in S$. Therefore, $ \dln p^\prime(\bar{a})\in
S(r,r')$. But this implies that $r'\not\models p(\bar{a})$. Thus,
$p(\bar{a})\in\Delta(r,r')$.

\item[(II)] $r\not\models p(\bar{a})$ and $I(S)\models p(\bar{a})$. In
this case, $p(\bar{a})\not\in S$ ($S$ is a minimal model and
$p(a)$ does not need to be in $S$ if it was not in $r$). Thus, by
definition of $I(S)$ we conclude that $ p^\prime(\bar{a})\in S$.
Therefore, $ p^\prime(\bar{a})\in S(r,r')$. But this implies that
$r'\models p(\bar{a})$. Thus, $p(\bar{a})\in\Delta(r,r')$.
\end{enumerate}
Thus, $\Delta(r,I(S))\subseteq\Delta(r,r')$. But, by proposition
\ref{l1}, $I(S)$ satisfies $IC$, and therefore, $\Delta(r,I(S))$
must be equal to $\Delta(r,r')$, since $\Delta(r,r')$ is minimal
under set inclusion in $\{ \Delta(r,r^*) \mid r^* \models IC \}$.
Then, we conclude that $I(S)=r'$. \boxtheorem\\

\Proof{Theorem \ref{tw}}  We will prove the first part of this
theorem. The second one can be proved analogously.

Given a repair $r'$ of $r$, by lemma \ref{l3}, $r'=I(S_M)$, where
$S_M$ is an answer set of $\Pi_\Delta(r)$, with $S_M\subseteq
S(r,r')$. Define $S$ from $S_M$ as in (\ref{e2}). We obtain that
$S$ is an answer set of $\Pi(r)$. By construction of $S$, $I(S) =
I(S_M)$. Furthermore, $I(S) = \{ p(a) ~|~ p'(a) \in S\}$.
\boxtheorem\\

\Proof{Proposition \ref{prop:wfs}} Since it is always the case
that $\w \subseteqq \C$ \cite{infComp}, we only need to show that
$\C \subseteqq \w$. In consequence, it is necessary to check that
whenever a literal $(\neg)p'(a)$ belongs to $\C$, where $a$ is
tuple of elements in the domain $D$ and $p$ is a database
predicate, $(\neg)p'(a)$ can be fetched into $\W^n(\emptyset)$ for
some finite integer $n$.

Each literal $L$ in the original database $r$, in its primed
version, will become $L$ or its complement $\overline{L}$ in the
answer sets\footnote{Actually only positive literals appear in
$r$, but we are invoking the CWA. All the literals in the original
instance will belong to $\nit{Core}(\Pi(r))$.}. We will do the
proof by cases, considering for a literal  $L:~ (\neg) p'(a)$
contained in $\nit{Core}(\Pi(r))$ all the possible transitions
from the original instance to the core: (a) negative to positive.
(b) positive to positive. (c) negative  to negative. (d) positive
to negative. We will prove only the first two cases, the other two
are similar.

For each  case, again several cases have to be verified according
to the different ground program rules that could have made $p'(a)$
get into $\C$.\\
\\
(I) Assume $p'(a) \in \C$. To prove: $p'(a) \in \w$.  Two cases

\begin{enumerate}
\item  $p(a) \notin r$. Since FDs can only produce deletions
$p'(a)$ has to be true because an unary constraint was false for
$p(a)$: ~$(p(a) \vee \varphi(a)) \in IC_D$ is false, where $IC_D$
is the instantiation of the ICs in the domain $D$. Then,
$\varphi(a)$ is false. In the ground program we find the rule~
$p'(a) \lla dom(a), \neg \varphi(a)$. The second subgoal becomes
true of $\emptyset$. Since $dom(a) \in \W^1(\emptyset)$. Then,
$p'(a) \in \W^2(\emptyset)$.

\item $p(a) \in r$. Intuitively, $p(a)$ persisted. This means there is
no ground IC of the form $p'(a) \vee \varphi(a)$ that is false,
nor of the form $\neg p'(a) \vee \psi(a)$, with $\psi(a)$ false.
Otherwise, $p'(a)$ could not be in the core. The second case
implies that we can never obtain $\neg p'(a)$ by means of a rule
of the form ~$\neg p'(a) \lla dom(a), \neg \psi(a)$. Some cases
need to be examined:

\begin{enumerate}
\item There is $(p(a) \vee \varphi(a))$ with $\varphi(a) \in IC_D$ false. Then, with the
rule ~$p'(a) \lla dom(a),\neg \varphi(a)$, with $\varphi(a)$. In
this case, as in case 1., $p'(a) \in \W^2(\emptyset)$.

\item If there is no ground constraint as in the previous item,
either because there is no  $(p(a) \vee \varphi(a))$ in $IC_D$ or
the $\varphi(a)$s are true, then there is no applicable rule of
the form ~$p'(a) \lla dom(a),\neg \varphi(a)$ in the ground
program. Since rules associated to FDs delete tuples only, we
could obtain $p'(a)$  due to a default rule $p'(a) \lla dom(a),
p(a), not~ \neg p'(a)$ only,  via the unfoundedness of $\neg
p'(a)$, or directly via the unfoundedness of $\neg p'(a)$ in the
ground program. If the $\W$ operator declares $\neg p'(a)$
unfounded, then $p'(a)$ will belong to $\w$. So, we have to
concentrate on the unfoundeness of $\neg p'(a)$.

\begin{enumerate}

\item We know that we can never get $\neg p'(a)$ from rules of the
form $\neg p'(a) \lla dom(a), \neg \psi(a)$.

\item $\neg p'(a)$ cannot be obtained via the default rule,
because $\neg p'(a) \lla dom(a), not~ p(a), not~ p'(a)$ has the
second subgoal false.

\item $\neg p'(a)$ cannot be obtained via a possible unfoundedness
of $p'(a)$, because $p'(a)$ belongs to answer sets.

\item We are left with rules associated to FDs. Assume that
$(\neg p(a) \vee \neg p(b) \vee c = d) \in IC_D$. There are only
two alternatives: $c =d$, in which case, the associated triggering
rule cannot be applied; or $c \neq d$ and $\neg p(b)$ is true and
there is no $(p(b) \vee \chi(b)) \in IC_D$ with $\chi(b)$ false
(otherwise, $\neg p'(a)$ would have to be true).

The rule ~$\neg p'(a) \vee \neg p'(b) \lla dom(a), dom(b), p(a),
p(b), c \neq d$ cannot be applied, because $p(b)$ is false.

We have to analyze the stabilizing rule ~$\neg p'(a) \lla  p'(b),
c \neq d$. If $c =d$, the rule does not apply. Otherwise,  we have
$p(b) \notin r$, and then $\neg p(b) \in \W^(\emptyset)$. $p'(b)$
cannot be obtained from the default $p'(b) \lla dom(b), p(b), not~
\neg p'(b)$, because $p(b)$ is false. Neither can it be obtained
from a rule $p'(b) \lla dom(b), \neg \chi(b)$, because $\chi(b)$
would have to be true.

In consequence, $p'(b)$ is unfounded, i.e. $\neg p'(b) \in
\W^2(\emptyset)$, then $\neg p'(a)$ turns out to be unfounded:
 $p'(a) \in \W^3(\emptyset)$.
\end{enumerate}
\end{enumerate}
\end{enumerate}

\noindent (II) $\neg p'(a) \in \C$. The other (similar) cases are:
\begin{enumerate}
\item $p(a) \notin r$ and $\neg p'(a) \in \C$.

\item $p(a) \in r$ and $\neg p'(a) \in C$.

\end{enumerate}

It is possible to show that always $\C \subseteq \W^3(\emptyset)$.
\boxtheorem\\


\begin{thebibliography}{}
\bibitem[\protect\citename{Abiteboul {\em et al.}, }1995]{AbHuVi95}
Abiteboul, S.; Hull, R.; and Vianu, V.
\newblock {\em Foundations of
Databases}. \newblock Addison-Wesley, 1995.


\bibitem[\protect\citename{Arenas \emph{et al.}, }1999]{pods99}
Arenas, M.; Bertossi, L.; and Chomicki, J.
\newblock {Consistent Query Answers in Inconsistent Databases}.
\newblock In {\em Proc. ACM Symposium on Principles of Database Systems (ACM
  PODS'99, Philadelphia)}, ACM Press, 1999, pp. 68--79.

\bibitem[\protect\citename{Arenas \emph{et al.}, }2000a]{fqas2k}
Arenas, M.; Bertossi, L.; and Chomicki, J.
\newblock {Specifying and Querying Database Repairs using Logic Programs with
  Exceptions}.
\newblock In
{\em Flexible Query Answering Systems. Recent Developments}, H.L.
Larsen, J. Kacprzyk, S. Zadrozny, H. Christiansen (eds.),
Springer, 2000, pp. 27--41.


\bibitem[\protect\citename{Arenas \emph{et al.}, }2001]{icdt01}
Arenas, M.; Bertossi, L.; and Chomicki, J.
\newblock {Scalar Aggregation in FD-Inconsistent Databases}.
\newblock In {\em Database Theory - ICDT 2001} (Proc. International Conference on Database
Theory,
  ICDT'2001), Springer, Lecture Notes in Computer Science 1973,
  2001,
  pp. 39 -- 53.

\bibitem[\protect\citename{Arenas \emph{et al.}, }2000b]{annotated}
Arenas, M.; Bertossi, L.; and Kifer, M.
\newblock {Applications of Annotated Predicate Calculus to Querying
  Inconsistent Databases}.
\newblock In {\em `Computational Logic - CL 2000'. Stream: 6th International
  Conference on Rules and Objects in Databases (DOOD'2000)},
  Springer, Lecture Notes in Artificial Intelligence 1861, 2000, pp. 926--941.


\bibitem[\protect\citename{Buccafurri \emph{et al.}, }2000]{weak}
Buccafurri, F.; Leone, N.; Rullo, P.
\newblock {Enhancing Disjunctive Datalog by Constraints}.
\newblock {\em IEEE Transactions on Knowledge and Data Engineering}, 2000, 12(5): 845-860.

\bibitem[\protect\citename{Celle and Bertossi, }2000]{celle-bertossi2k}
Celle, A.; and Bertossi, L.
\newblock {Querying Inconsistent Databases: Algorithms and Implementation}.
\newblock In {\em `Computational Logic - CL 2000'. Stream: 6th International
  Conference on Rules and Objects in Databases (DOOD'2000)}, pp. 942--956.
  Springer Lecture Notes in Artificial Intelligence 1861.

\bibitem[\protect\citename{Chou and Winslett, }1994]{winslett94}
Chou, T.; and Winslett, M.
\newblock {A Model-Based Belief Revision System}.
\newblock {\em Journal of Automated Reasoning}, 1994, 12:157--208.

\ignore{
\bibitem[\protect\citename{Citrigno \emph{et al.}, }1997]{dlvSys}
Citrigno, S. et al.
\newblock {The DLV System: Model
       Generator and Advanced Frontends (system description)}.
\newblock In {\em Proc.
Twelfth Workshop Logic Programming (WLP'97)}. F. Bry, B. Freitag,
D. Seipel (eds.). Technical Report PMS-FB-1997-10 of the Ludwig
Maximilians Universitaet M\"{u}nchen, 1997. }

\bibitem[\protect\citename{Dalal, }1988]{dalal}
Dalal, M.
\newblock {Investigations into a Theory of Knowledge Base
Revision: preliminary report}.
\newblock In {\em Proc. Seventh National Conference on Artificial
Intelligence (AAAI'88)}, 1988, pp. 475--479.

\bibitem[\protect\citename{Dantsin {\em et al.}, }200?]{voronkov}
Dantsin, E., Eiter, T., Gottlob, G. and Voronkov, A.
\newblock Complexity and Expressive Power of Logic Programming.
\newblock To appear in ACM Computing Surveys.

\bibitem[\protect\citename{Eiter \emph{et al.}, }1998]{E*98}
Eiter, T.; Leone, N.; Mateis, C.; Pfeifer, G.; and Scarcello, F.
\newblock {The Knowledge Representation System DLV: Progress Report,
  Comparisons, and Benchmarks}.
\newblock In {\em Proceedings of the International Conference on Principles of
  Knowledge Representation and Reasoning, KR98, Trento, Italy, June 1998}.
  Morgan Kaufman.

\bibitem[\protect\citename{Eiter \emph{et al.}, }2000]{dlvSys}
Eiter, T.; Faber, W.; Leone, N.; Pfeifer, G.
\newblock {Declarative Problem-Solving in DLV}.
\newblock In {\em Logic-Based Artificial Intelligence}, J. Minker
(ed.), Kluwer, 2000, pp. 79--103.


\bibitem[\protect\citename{Fitting, }1996]{Fitting:96}
Fitting, M.
\newblock {\em First Order Logic and Automated Theorem Proving}.
\newblock Texts and Monographs in Computer Science. Springer Verlag, 2nd ed., 1996.

\bibitem[\protect\citename{Gelfond and Lifschitz, }1988]{gl88}
Gelfond, M.; and Lifschitz, V.
\newblock {The Stable Model Semantics for Logic Programming}.
\newblock In      {\em Logic Programming, Proceedings of the Fifth International
Conference and Symposium},
       R. A. Kowalski and  K. A. Bowen (eds.), MIT Press 1988,
pp. 1070--1080.

\bibitem[\protect\citename{Gelfond and Lifschitz, }1991]{glELPb}
Gelfond, M.; and Lifschitz, V.
\newblock {Classical Negation in Logic Programs and Disjunctive Databases}.
\newblock {\em New Generation Computing}, 1991, 9:365--385.



\bibitem[\protect\citename{Greco and Zumpano, }2000]{greco1}
Greco, S.; and Zumpano, E.
\newblock Querying Inconsistent Databases.
\newblock In Proc. 7th International Conference on Logic for Programming
and Automated Reasoning (LPAR'2000), Springer LNCS  1955,  2000,
pp.  308-325.

\bibitem[\protect\citename{Greco and Zumpano, }2001]{greco2}
Greco, S.; and Zumpano, E.
\newblock Computing Repairs for Inconsistent Databases.
\newblock Proc.  The Third International Symposium on
  Cooperative Database Systems for Advanced Applications
  (CODAS01), Beijing, April 23-24, 2001.

\bibitem[\protect\citename{Greco {\em et al.}, }2001]{greco3}
Greco, G.; Greco, S.; and Zumpano, E.
\newblock A Logic Programming Approach to the Integration,
Repairing and Querying of  Inconsistent Databases.
\newblock In  Proc. 17th International
Conference on Logic Programming, ICLP'01, Ph. Codognet (ed.), LNCS
2237, Springer, 2001, pp. 348--364.

\bibitem[\protect\citename{Kowalski and Sadri, }1991]{ks91}
Kowalski, R.; and Sadri, F.
\newblock {Logic Programs with Exceptions}.
\newblock {\em New Generation Computing}, 1991, 9:387--400.

\bibitem[\protect\citename{Leone \emph{et al.}, }1997]{infComp}
Leone, N.; Rullo, P.; and Scarcello, F.
\newblock {Disjunctive Stable Models: Unfounded Sets, Fixpoint
Semantics, and Computation}.
\newblock {\em Information and Computation}, 1997, 135(2):69-112.

\bibitem[\protect\citename{Leone, }2000]{pers}
Leone, N.
\newblock Personal communication, 2000.

\bibitem[\protect\citename{Lifschitz and Turner, }1994]{lifsTur94}
Lifschitz, V. and Turner, H.
\newblock Splitting a Logic
Program. \newblock
In Proceedings of the Eleventh International
Conference on Logic Programming, Pascal van Hentenryck (ed.), MIT
Press, 1994, pp. 23-37.

\bibitem[\protect\citename{Lloyd, }1987]{lloyd87}
Lloyd, J.W.
\newblock {\em {F}oundations of {L}ogic {P}rogramming}.
\newblock Springer Verlag, 1987.

\bibitem[\protect\citename{Marek and Truszczynski, }1998]{mt}
Marek, V. W.; and Truszczynski, M.
\newblock {Revision Programming}.
\newblock {\em Theoretical Computer Science}, 1998, 190(2):241--277.

\bibitem[\protect\citename{Ullman, }1988]{ullmanI}
 Ullman, J.
\newblock {\em Principles of {D}atabase and {K}nowledge-{B}ase Systems, {V}ol.
{I}}.
\newblock Computer Science Press, 1988.

\bibitem[\protect\citename{Winslett, }1988]{winslett88}
Winslett, M.
\newblock {Reasoning about Action using a Possible Model
Approach}.
\newblock In {\em Proc. Seventh National Conference on Artificial
Intelligence (AAAI'88)}, 1988, pp. 89--93.


\end{thebibliography}
\end{document}